\documentclass[lettersize,journal]{IEEEtran}
\usepackage{amsmath,amsfonts}
 \usepackage{array}
 \usepackage{stfloats}
\usepackage{url}
\usepackage{verbatim}
 \usepackage[noend]{algorithmic}
\usepackage[vlined,ruled,commentsnumbered,linesnumbered]{algorithm2e}
\usepackage{tabularx}
\usepackage{multirow}
\usepackage{sidecap}
\usepackage{subfig}
\usepackage{subfloat}
\usepackage{graphicx}
\usepackage{textcomp}
\usepackage{booktabs} 
\usepackage{pifont}
\usepackage{lipsum}
\usepackage[outline]{contour}
\usepackage{subfloat}
\usepackage{comment}

\usepackage[utf8]{inputenc}

\usepackage{filecontents}
\newcommand{\sys}{\texttt{ICMarks}}
\newcommand{\baseconstcap}{Cell Scattering}%
\newcommand{\baseinvasivecap}{Buffer Insertion}%
\newcommand\revision[1]{\textcolor{black}{#1}}
\newcommand\revisionsecond[1]{\textcolor{black}{#1}}

\usepackage{tikz}

\newcommand*\emptycirc[1][1ex]{\tikz\draw (0,0) circle (#1);} 
\newcommand*\halfcirc[1][1ex]{%
  \begin{tikzpicture}
  \draw[fill] (0,0)-- (90:#1) arc (90:270:#1) -- cycle ;
  \draw (0,0) circle (#1);
  \end{tikzpicture}}
\newcommand*\fullcirc[1][1ex]{\tikz\fill (0,0) circle (#1);} 
\DeclareFontEncoding{LS1}{}{}
\DeclareFontSubstitution{LS1}{stix}{m}{n}
\DeclareSymbolFont{symbols4}{LS1}{stixbb}{m}{it}
\DeclareMathSymbol{\varhexagonblack}{\mathord}{symbols4}{"DD}
\DeclareMathSymbol{\hexagonblack}   {\mathord}{symbols4}{"DE}
\definecolor{gray}{gray}{0.7}
\newcommand{\redborderhexagon}{
    \textcolor{red}{\scalebox{1.3}{$\varhexagonblack$}}%
    \llap{\textcolor[HTML]{BFA100}{$\varhexagonblack$}}
}
\def\BibTeX{{\rm B\kern-.05em{\sc i\kern-.025em b}\kern-.08em
    T\kern-.1667em\lower.7ex\hbox{E}\kern-.125emX}}
\usepackage{balance}
\begin{document}
\title{\sys: A Robust Watermarking Framework for Integrated Circuit Physical Design IP Protection}
\author{Ruisi Zhang, Rachel Selina Rajarathnam, David Z. Pan, Farinaz Koushanfar 
\thanks{Ruisi Zhang and Farinaz Koushanfar are with the Department of Electrical and Computer Engineering, University of California, San Diego, CA. (e-mail: ruz032@ucsd.edu; fkoushanfar@ucsd.edu)\\
Rachel Selina Rajarathnam and David Z. Pan are with the Department of Electrical and Computer Engineering, The University of Texas at Austin, Austin, TX. (e-mail: rachelselina.r@utexas.edu; dpan@ece.utexas.edu)
}}
\IEEEaftertitletext{\vspace{-2.2\baselineskip}}
\markboth{Journal of \LaTeX\ Class Files,~Vol.~18, No.~9, September~2020}%
{How to Use the IEEEtran \LaTeX \ Templates}

\maketitle

\begin{abstract}
Physical design watermarking on contemporary integrated circuit (IC) layout encodes signatures without considering the dense connections and design constraints, which could lead to performance degradation on the watermarked products. 
This paper presents \sys, a quality-preserving and robust watermarking framework for modern IC physical design. 
\sys{} embeds unique watermark signatures during the physical design's placement stage, thereby authenticating the IC layout ownership.
\sys's novelty lies in (i) strategically identifying a region of cells to watermark with minimal impact on the layout performance and (ii) a two-level watermarking framework for augmented robustness toward potential removal and forging attacks. 
Extensive evaluations on benchmarks of different design objectives and sizes validate that \sys{} incurs no wirelength and timing metrics degradation, while successfully proving ownership. 
Furthermore, we demonstrate \sys{} is robust against two major watermarking attack categories, namely, watermark removal and forging attacks; even if the adversaries have prior knowledge of the watermarking schemes, the signatures cannot be removed without significantly undermining the layout quality.

\end{abstract}


\section{Introduction}
\label{sec:intro}
In the modern very-large-scale integrated circuit (VLSI) supply chain, companies across multiple countries collaborate to design, fabricate, assemble, and verify integrated circuits (ICs)~\cite{liu2016vlsi,pawar2018risks}.
The final product integrates intellectual property (IP) components from various stakeholders, including design~\cite{5938070,divyanshu2023fsm,4415879,sun2006watermarking} and fabrication houses~\cite{saha2014watermarking}, along the global supply chain.
With the confluence of diverse inputs, safeguarding IP emerges as an undeniable critical necessity, particularly to preempt IP piracy threats~\cite{shamsi2019ip,knechtel2019protect,slpsk2023treehouse}.
The physical design of the supply chain bridges the logic design with the VSLI layout for manufacturing. It strategically optimizes the positioning and connectivity of components on the chip canvas and enhances power-performance-area (PPA) metrics. 
The optimizations not only translate into million-dollar manufacturing cost savings for chip producers~\cite{sutardja20151} but also establish invaluable IPs. However, they are vulnerable to unauthorized use, like forging attacks by malicious third parties in the supply chain~\cite{6860363}, and require robust protective measures.


Watermarking~\cite{rathor2023hard,chen2023intellectual,tauhid2023survey} has emerged as a viable methodology to safeguard physical design IP by embedding unique and confidential signatures into the IC layout.
To authenticate ownership, the design houses verify the watermark signatures by decoding the watermarks from the IC layout. Existing physical design watermarking frameworks~\cite{6860363} proposed to insert watermarks from two directions: (i) constraint-based  watermarking~\cite{saha2007novel,bai2007watermarking,ni2004watermarking,kahng2001constraint,kahng1998robust}, and, (ii) invasive watermarking~\cite{nie2005watermarking,sun2006watermarking}. Constraint-based watermarking encodes signatures by setting cell position/topology as optimization constraints during the physical design.
Such frameworks, primarily designed for partition-based placement algorithms~\cite{dunlop1985procedure,tsay1991unified}, do not consider new design constraints in modern IC layouts, such as fence regions, leading to quality degradation. 
The invasive watermarking adds redundant cells~\cite{sun2006watermarking} or wires~\cite{nie2005watermarking} as watermarks. However, the signatures could \revision{be forged~\cite{guin2014counterfeit}} if the adversary has knowledge of the watermarking algorithms.




Inserting watermarks into modern IC layouts is non-trivial and exhibits several challenges. First, large-scale designs have dense standard cell connections, and slight perturbations of the cell locations/topologies can degrade the overall performance. Second, modern ICs often have additional design constraints like fence regions and macros. Developing watermarking algorithms without considering these additional design constraints will negatively affect overall performance on certain layouts. Finally, the watermarks should be robust against potential removal and forging attacks from the supply chain.


This paper devises \sys{}, a novel and robust watermarking framework to safeguard IC layout IP.
It consists of two consecutive steps, namely, \textit{Global Watermarking} (GW) and \textit{Detailed Watermarking} (DW), and targets the watermark insertion at the global and detailed placement stages, respectively. 
\textit{Global Watermarking} (GW) leverages a novel scoring scheme to identify a region to watermark that does not violate the design constraints and ensures minimal performance deterioration. The identified watermarked region, along with the associated cells inside it, encodes the GW signature by co-optimizing the placement such that only the associated cells intersect the watermarked region.  Next, \textit{Detailed Watermarking} (DW) selects cells that do not overlap with nearby cells when moving along the x/y axis within GW's watermarked region. 
The cells are watermarked by shifting in the x/y directions to encode DW signature in the detailed placement.
As such, the strategically inserted watermarked region and cells ensure the signature insertion satisfies the modern IC design constraints and incurs minimal quality degradation.
The design company proves ownership by reverse engineering~\cite{9300272,alrahis2021gnn} the logic netlist and all standard cell locations from an IC layout in GDSII format, and employ \sys{} to decode signature for ownership verification.

\sys{} is robust against threats from the supply chain that aim to remove or forge the watermarks. 
By encoding signatures as placement co-optimization objectives, it can withstand a spectrum of watermark removal attacks~\cite{6860363}. Extensive targeted watermark removal attacks in different layout regions validate that \sys{} still maintains over 90\% watermark extraction rates when the adversaries compromise a maximum of $\sim$2\% wirelength/timing quality.  In addition, the two-level augmented GW and DW signatures make the watermark strength stronger compared with the best prior practice~\cite{kahng1998robust,kahng2001constraint}, and forging the signatures becomes exceedingly harder.  \revision{The watermarking strength is measured by the probability that the watermark constraints are satisfied by coincidence.} 

In brief, our contributions are as follows:

\begin{itemize}
\item  We present \sys, the first watermarking scheme with both region and position constraints for the modern VLSI layouts, which does not degrade the layout quality while being robust against watermark removal and forging attacks. 
\item  Our framework features: (i) a novel search algorithm to identify the watermarked region that adheres to design constraints and incurs minimal performance degradation in \textit{Global Watermarking}; (ii) an innovative encoding mechanism for quality-preserving signature insertion and augmented robustness in \textit{Detailed Watermarking}.
\item  We conduct experiments on benchmarks of different design objectives and sizes: (i) \sys{} introduces no degradations on the wirelength-driven ISPD'2015~\cite{bustany2015ispd} and ISPD'2019~\cite{liu2019ispd} benchmarks, whereas the best prior method~\cite{4415879} degrades the average routed wirelength by up to 1.4\%; (ii) \sys{} shows no degradations on the timing-driven ICCAD'2015~\cite{7372671} benchmarks, whereas the best prior method~\cite{kahng1998robust,kahng2001constraint} degrades the total and worst negative slack by 1.14\% and 1.51\% respectively~\footnote{The IC physical design is fragile that 0.5\% performance degradation can compromise the additional efforts from the design company for quality metrics optimization~\cite{qiu2023progress}.}. 
\item  We perform comprehensive robustness analysis: (i) \sys{} withstands watermark removal attacks targeting different layout regions and maintains over 90\% watermark extraction rates; (ii) \sys{} resists watermark forging attacks, making it hard to counterfeit the signatures.

\end{itemize}

\section{Background and Related Work}
\label{sec:background}
In this section, we introduce the background for VLSI placement and IC design watermarking. 

\subsection{VLSI Placement}

\begin{figure}[ht]
    \centering
    \includegraphics[width= \columnwidth]{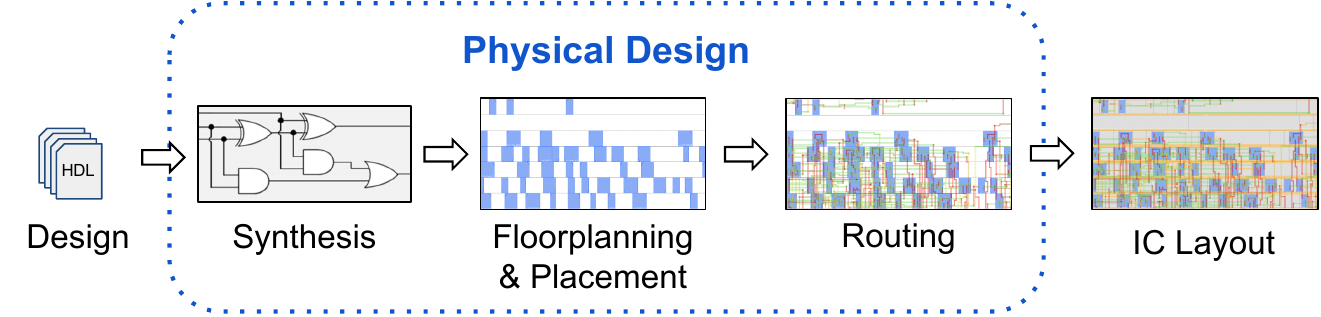}
    \caption{The VLSI physical design process. It generates the IC layout from a design RTL through synthesis, placement, and routing stages.}
    \label{fig:physical_design}
\end{figure}

The integrated circuit (IC) design flow starts with design specifications that are converted to register transfer level (RTL) using a hardware description language (HDL). As depicted in Fig.~\ref{fig:physical_design}, the physical design process synthesizes the design HDL to generate a logic-level netlist comprising macros and standard cells from the technology library. 
A floorplanning step determines regions and locations for the design's macros and the IO (input-output) ports.
Typically, the standard cells are placed during the placement stage, followed by routing all the design connectivity.
The routed design is verified to ensure manufacturability and the physical layout is generated in GDSII format to send to the fabrication unit for manufacturing.
 
With fixed macro locations,
placement is an essential step in the physical design flow to determine the locations of all the standard cells in the design~\cite{qiu2023progress,pawanekar2015nonlinear,agnesina2020vlsi,lin2021routability}. 
It significantly impacts the quality and efficiency of the subsequent stages, 
including routing and manufacturing.
The VLSI placement typically consists of three stages: Global Placement, Legalization, and Detailed Placement.

\subsubsection{Global Placement} 
In the global placement stage, the placer aims to achieve roughly legal cell locations by distributing the cells across the chip area to minimize overlaps.
In addition, the placer also targets other objectives, such as minimizing wirelength and ensuring timing constraints are met.
 
\noindent\textbf{Wirelength-driven Placement:} A wirelength-driven placer aims to minimize the overall wirelength $W$ while ensuring minimal cell overlap, as specified in Eqn.~\ref{eqn:general_place}~\cite{lin2019dreamplace}.
\begin{equation} \label{eqn:general_place}
    min_{x,y} W(x,y) \ s.t.\  D(x,y) < \mathcal{D}
\end{equation}
where $(x,y)$ is the location of cells in the placement, and $W(x,y)$ denotes the total wirelength \revision{connecting cells} in the design. $D(x,y)$ is a density measure of the overlaps among the cells, and it should be below the threshold $\mathcal{D}$.

\noindent\textbf{Timing-driven Placement:}
A timing-driven placer prioritizes optimizing timing critical paths in the design while ensuring the overlaps among the cells are minimum~\cite{9774725}.
Given a timing endpoint $t$ (a primary output port or an input pin of memory element) with an arrival time of $t_{\mathrm{AT}}(t)$ and a required arrival time $t_{\mathrm{RAT}}(p)$, the \textit{slack} of $t$ is calculated as follows,
\begin{equation} \label{eqn:timing_comp}
\begin{aligned}
slack(t) &= t_{\mathrm{RAT}}(t)-t_{\mathrm{AT}}(t) \\
TNS &= \sum_{t \in P_{\text {end }}} min(0,slack(t)) \\
WNS &= \min_{t \in P_{\text {end }}} slack(t) \\
\end{aligned}
\end{equation}
With the timing endpoints $P_{\text {end }}$ in the design, the total negative slack \textit{TNS} and the worst negative slack \textit{WNS} are computed as shown in Eqn.~\ref{eqn:timing_comp}.
A timing-driven placer optimizes either the \textit{WNS} or the \textit{TNS} to meet the timing constraint of the design.
 
\subsubsection{Legalization} The legalization stage 
moves cells to eliminate the cell overlaps, ensure the row alignments, and guarantee the the design constraints are met~\cite{spindler2008abacus,netto2021algorithm,yang2022mixed}.
The multi-row height cells are prioritized for legalization before the single-row height cells.




\subsubsection{Detailed Placement}
The detailed placement stage refines the legalized placement to improve design metrics such as wirelength or timing constraints~\cite{pan2005efficient,chow2014cell}.
The refinement can be achieved by operations like (i) Swapping the positions of two cells to improve the objectives without causing legality or timing violations and (ii) Window-based approaches to find optimal cell locations within the specified region.
 
\subsection{Watermarking in IC Design}\label{subsec:wm_chip}


Watermarking~\cite{rathor2023hard,chen2023intellectual,tauhid2023survey} encodes unique and confidential signatures into the IC layouts to assist product owners in proof of ownership.
Design houses leverage this technique to enhance IP rights protection in the chip supply chain and detect unauthorized usages or replications. 
These watermarks are typically inserted in the logic design or physical design level. 

\subsubsection{Logic Design Watermarking} 
The watermarking at the logic design level focuses on safeguarding the Register Transfer Level (RTL) or netlist ownership. Designers embed unique signatures into the RTL or netlist that do not impact the core functionality and are invisible upon adversaries' inspection. The signature insertion can be categorized as follows: (i) modifying the finite state machine (FSM) to add additional inputs or unused components as watermarks~\cite{5938070,divyanshu2023fsm}; (ii) adding additional power components as side-channel signatures~\cite{becker2010side,das2023psc}; and (iii) introducing unique triggers during netlist design that induces the malfunctions when the triggers are detected~\cite{8764416}.


More recent works introduce logic locking to add additional logic gates and circuits that alter the original design's functionality when the incorrect key is applied~\cite{sengupta2020truly,kamali2022advances}; obfuscation to adding dummy logic gates or additional circuit connections to make the logic design appear differently from the original one~\cite{santikellur2021hardware,bagul2023hardware}. These techniques are orthogonal to the logic design watermarking, which can be combined to establish more robust intellectual property protection (IPP) frameworks.

\revision{Logic design watermarking faces several limitations:
(i) Adding additional watermarks into the logic design needs to consider physical design criteria like timing to maintain chip quality. However, watermarking after these constraints have been optimized at the physical design stage reduces such discrepancies. (ii) In logic design, if the logic circuits are not performing correctly after watermark insertion, designers need to revisit the RTL design in the previous stage. Physical watermarking, however, does not modify the circuits' functionalities and only changes how cells are placed/connected on the canvas.
}

\subsubsection{Physical Design Watermarking} The watermarking at the physical design level encodes signatures onto the IC layout from two directions: (i) constraint-based  watermarking~\cite{saha2007novel,4415879,bai2007watermarking,ni2004watermarking,kahng2001constraint,kahng1998robust} that uses cell topology/position as the additional constraints during physical design. \revision{Row Parity~\cite{kahng1998robust,kahng2001constraint} constraints 1-bit watermark cells to be on the odd row and the 0-bit watermark cells to be on the even row. Cell Scattering~\cite{4415879} enforces the 1-bit watermark cell to move one unit along the y-axis and 0-bit cell one unit along the x-axis.} (ii) invasive watermarking~\cite{nie2005watermarking,sun2006watermarking} that adds redundant cells or wires as the watermark signatures.

The watermarking frameworks' quality is evaluated by the following metrics.

\noindent\textbf{Criteria 1 - Fidelity:} The watermarked layouts shall incur minimal quality degradation compared with non-watermarked ones on metrics like wirelength and timing. Besides, the watermarked layouts should adhere to the layouts' design constraints, such as fence regions and row alignments.

\noindent\textbf{Criteria 2 - Efficiency:} The watermarking framework shall be efficient with minimal overheads for watermark insertion.

\noindent\textbf{Criteria 3 - Robustness:} The watermarked layout shall be robust against various attacks targeting to remove or forge the signature.


\begin{table}[h]
\centering
 \resizebox{\columnwidth}{!}{
\begin{tabular}{ccccc}
\toprule
Stage & WM Method & Fidelity & Efficiency & Robustness \\ 
\midrule
\begin{tabular}[c]{@{}c@{}}Logic\\ Design\end{tabular} & \begin{tabular}[c]{@{}c@{}}Logic Design\\ WM~\cite{5938070,becker2010side,8764416}\end{tabular}& \halfcirc & \emptycirc & \halfcirc  \\\hline
\multirow{5}{*}{\begin{tabular}[c]{@{}c@{}}Physical\\ Design\end{tabular}}
&\begin{tabular}[c]{@{}c@{}}Invasive WM\\~\cite{nie2005watermarking,sun2006watermarking}\end{tabular} & \halfcirc & \fullcirc & \emptycirc \\ \cline{2-5}
& \begin{tabular}[c]{@{}c@{}} Constraint-based WM\\~\cite{kahng2001constraint,kahng1998robust,saha2007novel,bai2007watermarking,ni2004watermarking}\end{tabular} & \halfcirc & \fullcirc & \fullcirc \\ \cline{2-5}
&\sys{} & \fullcirc &  \fullcirc &  \fullcirc\\
\bottomrule
\end{tabular}
}
\caption{Comparison of IC watermarking (WM) frameworks. 
\emptycirc{} framework does not meet criteria; \halfcirc{} framework partially meets criteria; \fullcirc{} framework completely meets criteria.
\label{tab:compare}}
\end{table}


The constraint-based watermarking frameworks~\cite{kahng2001constraint,kahng1998robust,saha2007novel,bai2007watermarking,ni2004watermarking} encodes signatures during the placement stage of physical design.
The watermarks are embedded as part of the layout. By encoding the constraints into the layout, constraint-based watermarking are robust against removal or forging attacks. The overhead of watermark insertion remains negligible compared to the time required to optimize the IC layout.  However, the watermarking frameworks do not consider the new design constraints, such as macros and fence regions that could lead to performance degradation.




By adding redundant cells/wires, the invasive watermark~\cite{nie2005watermarking,sun2006watermarking} is efficient for watermark insertion, and the quality degradation depends on how the corresponding algorithm designs the insertion mechanism. 
Nevertheless, the encoded watermarks can be forged if the adversaries have knowledge of the watermarking algorithm.



\section{Problem Formulation}
\label{sec:goals}
In this section, we introduce \sys's goal and its threat model.


\textbf{Watermarking Goal}
In the physical design stage, the IC design company invests huge efforts in identifying and fine-tuning optimization objectives for better cell placement and net routing. The optimization and customization are infused into different levels of the final layout, including (but not limited to) blocks and cells, to boost IC performance and reduce manufacturing expenses. As such, the final IC layouts constitute invaluable intellectual property, emphasizing the need to safeguard it.

\begin{figure}[ht]
    \centering
    \includegraphics[width= \columnwidth]{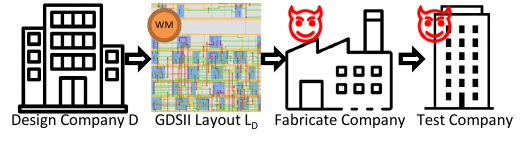}
     \caption{IC layout watermark scenario in the supply chain. The design company watermarks the IC layout  before manufacturing and testing.}
    \label{fig:scenario}
\end{figure}

\sys{} establishes ownership proof for the VLSI physical design by inserting watermarks into the placement stage, thereby authenticating the IP for the final layout. Fig.~\ref{fig:scenario} depicts a typical scenario within the supply chain. The IC design company $D$ uses highly customized and optimized placement and routing algorithms to enhance physical design quality. The layout $L_D$ from the physical design stage is watermarked and sent to the fabrication company for manufacturing. The manufactured ICs are verified for their functionalities in the test company. The design company $D$ can employ reverse engineering approaches~\cite{9300272,alrahis2021gnn} to extract the logic netlist and all standard cell locations from an IC layout in GDSII format, and employ \sys{} to decode the signature for ownership verification.

\textbf{Threat Model}\label{subsec:threat_model}
The adversary $\mathcal{A}$ in the fabricate or test company aims to steal the layout and stop $D$ from claiming its ownership by executing different attacks.

\noindent\textbf{Adversary’s Locations:} We consider the adversary $\mathcal{A}$ to be in the fabrication or test company and has access to the layouts produced by the design company.


\noindent\textbf{Adversary’s Objective:} Adversary $\mathcal{A}$'s objective is to (i) prevent design company $D$ from ownership proof by 
removing or forging the encoded signature; (ii) the attacked layout performance, such as wirelength or timing metrics shall not be significantly compromised. 


\noindent\textbf{Adversary’s Capacities:} We consider an adversary $\mathcal{A}$ with the following capacities: (i) The adversary $\mathcal{A}$ has access to the layout $L_D$. However, he cannot reverse-engineer the optimizations and customizations to reproduce $L_D$ because they are infused into every level of the design. (ii) The adversary $\mathcal{A}$ has access to open-source physical design tools to remove or forge the watermark. He also knows the general algorithms to watermark the layout. However, the random seeds, signatures, and insertion parameters are beyond his reach.

\section{\sys{} Design} 
\label{sec:method}

This section presents the proposed layout watermarking \sys{}, a two-level scheme that inserts watermarks during multiple stages of VLSI placement. In Section~\ref{sec:GW}, we introduce the \textit{Global Watermarking} method.  Next, we propose an independent embedding of the watermark signatures as \textit{Detailed Watermarking} in Section~\ref{sec:DW}. While \textit{Global} and \textit{Detailed Watermarking} can be used independently for IP protection, ICMarks combines both techniques to take advantage of their best properties.
 
\subsection{\textit{Global Watermarking} (GW)} \label{sec:GW}
\textit{Global Watermarking} embeds watermarks during the global placement stage. The watermarks, including a pre-defined watermarked region and associated cells in the region, are embedded as a co-optimization term during global placement. The co-optimization objective is to ensure only those associated cells are placed in the watermarked region. 

\subsubsection{\textbf{GW Watermark Selection}} 

\begin{figure}[ht]
    \centering
    \includegraphics[width= 0.95\columnwidth]{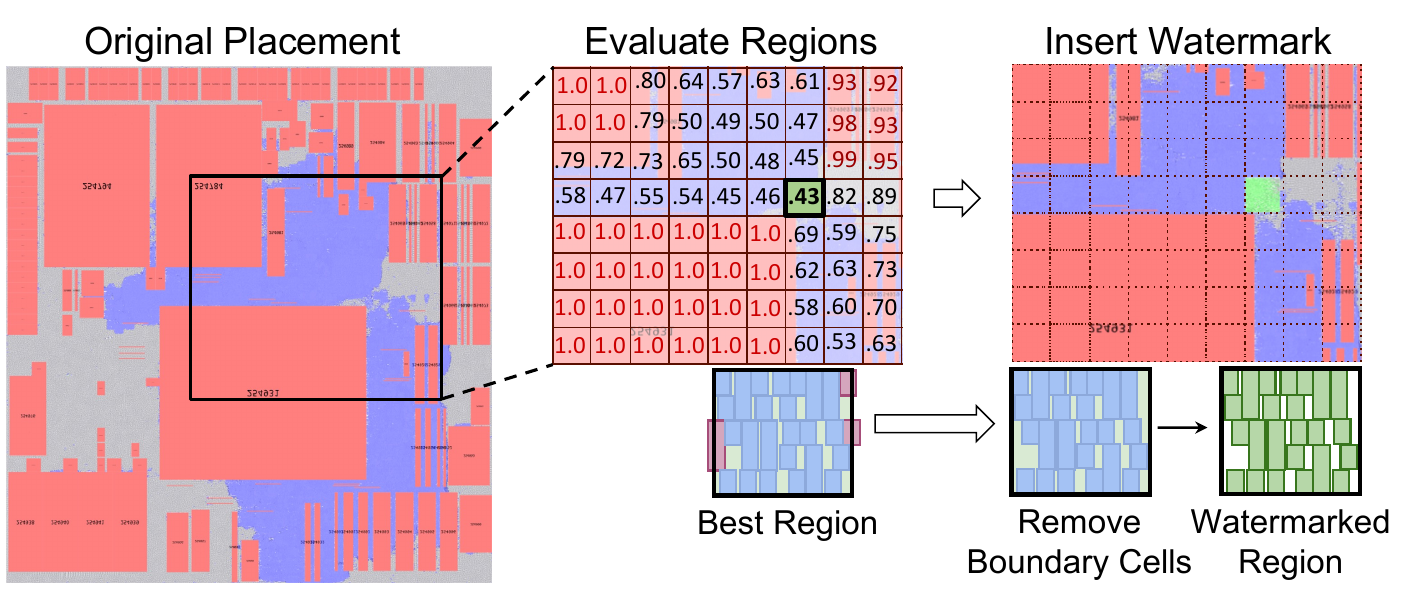}
     \caption{Global watermarking pipeline with stride = sliding window size.
     \revision{The \textcolor{blue}{blue} cells are the standard cells, the \textcolor{red}{red} cells are the macros, and the \textcolor{green}{green} cells are wm cells in GW.}}
    \label{fig:global}
\end{figure}

Given the original, optimized placement denoting the position of all cells as $P_{ori}$, we employ a sliding window-based algorithm that traverses the placement to search for a region to watermark as in Fig.~\ref{fig:global}.
An ideal watermarked region should meet three criteria: (i) \revision{as the signatures are encoded by moving cells along x/y-axis in DW,} the number of cells $N_c$ within the region should be sufficient to accommodate the required number of signature bits $N_w$; (ii) the total cell area $S_{cell}$ within the watermark region are $S$ shall be small, thereby providing ample space for watermarked cells to maneuver. \revision{The small watermark region utilization ensures the watermarks are encoded in the low-density regions.}; and (iii) the cells area $S_{overlap}$ overlap on the watermarked region boundary shall be minimized, such that their displacement from the watermarked region has minimal impact on the layout performance. These requirements are incorporated into an evaluation function $f$, as illustrated in Eqn.~\ref{eq:score}, to evaluate each region of the original placement $P_{ori}$. 
 
\begin{equation}
\label{eq:score}
    f = \alpha \frac{N_w}{N_c} + \beta \frac{S_{cell}}{S} + \gamma \frac{S_{overlap}}{S}
\end{equation}

The scoring function is balanced by adjusting the weights $\alpha$, $ \beta$, $\gamma$, where $\alpha, \beta, \gamma \in [0,1]$. The scores of the traversed regions are then normalized into a range of $[0, 1]$.
For regions that are nested within macros or other fence regions and any region where $N_c < N_w$, the evaluation function $f$ yields the highest evaluation scores as 1.0, rendering the regions not suitable for inserting GW watermarks. 
The region with the minimum score is chosen as our target watermarked region $R_w$, and the set of associated cells within the region is denoted as $C_{w1}$.

\subsubsection{\textbf{GW Watermark Insertion}}
The watermarked region $R_w$ and its associated cells $C_{w1}$ is formulated as the additional watermarked region constraint in the placement objective. It enforces the associated cells $C_{w1}$ to be in the watermarked region $R_w$ and other cells to be out of $R_w$.
For a design with $K$ fence regions, the placement formulation includes watermarked region $R_w$ as the additional optimization constraints, as specified in Eqn.~\ref{eq:multiregion}.

\begin{equation}
\label{eq:multiregion}
\begin{array}{ll}
\min _v & \sum_{e \in E} W(e ; v)+\lambda D(v), \\
\text { s.t. } & v_k=\left(x_k, y_k\right) \in R_k, \quad k=0, \cdots, R_K, R_w,
\end{array}
\end{equation}

$W$ is the wirelength term and $D$ is the cell density term with density multiplier $\lambda$. $v$ denotes the $(x,y)$ location of cell and $e\in E$ is the design net.
The watermarked region $R_w$ is added as an additional region constraint with associated cells $C_{w1}$ to obtain a watermarked placement $P_w$.

\subsubsection{\textbf{GW Watermark Extraction}}

The design owner asserts ownership by employing the watermarked region $R_w$ to extract cells $C_{w1}^\prime$ within the region and comparing these with the associated cells $C_{w1}$ and non-associated cells $C_{w1o}$. The extraction rate of the watermark, denoted as $WER_{GW}$, is calculated in Eqn.~\ref{eq:gw_wer}.

\begin{equation}
\label{eq:gw_wer}
\begin{array}{ll}
\%WER_{GW} = 100\times\frac{ |(C_{w1}^\prime \cap C_{w1}) - (C_{w1}^\prime \cap C_{w1o})|}{|C_{w1}|}
\end{array}
\end{equation}

\subsection{\textit{Detailed Watermarking} (DW)}\label{sec:DW}
In \textit{Detailed Watermarking}, the watermarked cells are embedded by moving them slightly along the x- or y-axis after legalization. A follow-up detailed placement stage compensates for the performance degradation from the watermark insertion. 

\subsubsection{\textbf{DW Watermark Selection}}
\begin{figure}[ht]
    \centering
    \includegraphics[width= 0.95\columnwidth]{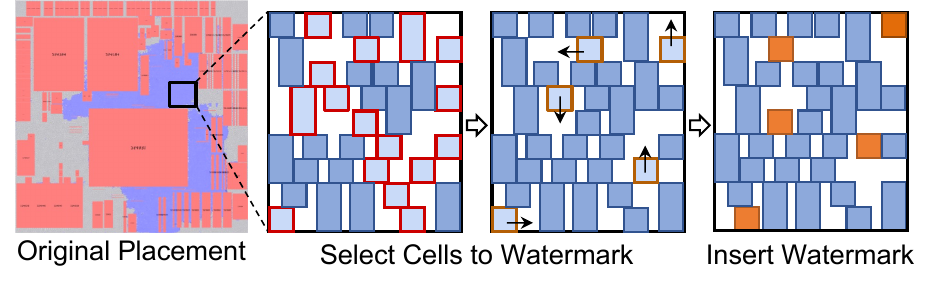}
     \caption{Detailed watermarking pipeline with \textcolor{purple}{Candidate cells} from which \textcolor{orange}{Watermarked cells} are selected.}
    \label{fig:detail_wm}
\end{figure}

The \textit{Detailed Watermarking} moves cells along the x- or y-axis to insert watermarks into the design. Randomly selecting watermarked cells and moving them without considering the dense interconnection could lead to significant performance degradation. To avoid this, \textit{Detailed Watermarking} only selects cells that will not overlap with their neighbors after cell movements as the watermark as in Fig.~\ref{fig:detail_wm}. 

Algorithm~\ref{alg:select_cells} outlines how \textit{Detailed Watermarking} identifies cells that will not overlap with neighbors during watermarking movements and subsequently inserts the signature. It starts by comparing the positions of cells within each row and identifying the cell indices that can move $|d_x|$ along the x-axis. Then, the algorithm compares cells across adjacent rows to determine if they can be moved up or down for $|d_y|$. The cells $C_x$ with room along the x-axis and $C_y$ with room along the y-axis are identified as candidate cells for watermarking. $D_x$ and $D_y$ are associated with these candidate cells to retain a record of their movement directions and distances. 

\setlength{\textfloatsep}{10pt} 
\begin{algorithm}[ht]
\caption{DW Watermark Selection and Insertion}
\label{alg:select_cells}
\small
\begin{algorithmic}[1]
    \REQUIRE Legalized placement \( (X_{lg}, Y_{lg}) \), Signature \( B_N \), Unit distance $|d_x|$ and $|d_y|$
    \ENSURE Watermarked cells \( C_{w2} \), Placement \((X_{wm}, Y_{wm})\)

    \STATE Initialize empty sets \( C_x, C_y, D_x, D_y \)
    
    \FOR{\( r = 0 \) to total\_num\_rows}
        \FOR{\( c \) in \( \text{cells\_in\_row}(r) \)}
            \IF{\( \text{is\_movable\_x}(c) \)}
                \STATE Append \( c \) to \( C_x \)
                \STATE \( s = \text{get\_direction\_x}(c) \)  \DontPrintSemicolon \tcp{\footnotesize{left: -1; right: 1}}
                \STATE Add \( (s \times |d_x|) \) to \( D_x \)
            \ENDIF
            \IF{\( \text{is\_movable\_y}(c) \)}
                \STATE Add \( c \) to \( C_y \)
                \STATE \( s = \text{get\_direction\_y}(c) \)  \DontPrintSemicolon \tcp{\footnotesize{down: -1; up: 1}}
                \STATE Add \( (s \times |d_y|) \) to \( D_y \)
            \ENDIF
        \ENDFOR
    \ENDFOR
    
    \STATE Random shuffle \( C_x \) and \( C_y \)
    \STATE \( X_{itr}, Y_{itr} = X_{lg}, Y_{lg} \)
    
    \FOR{\( i = 0 \) to \( |B_N| \)}
        \IF{\( B_N[i] == 1 \)}
            \STATE \( c_i = \text{random\_choose}(C_x) \)  \DontPrintSemicolon \tcp{\footnotesize{no replacement}}
            \STATE \( X_{itr}[c_i] = X_{lg}[c_i] + D_x[c_i] \)
        \ELSE
            \STATE \( c_i = \text{random\_choose}(C_y) \) \DontPrintSemicolon \tcp{\footnotesize{no replacement}}
            \STATE \( Y_{itr}[c_i] = Y_{lg}[c_i] + D_y[c_i] \)
        \ENDIF
        \STATE Add \( c_i \) to \( C_{w2} \)
    \ENDFOR
    
    \STATE \( X_{wm}, Y_{wm} = \textit{detailed\_placement}(X_{itr}, Y_{itr}) \)
    \STATE \textbf{return} \( C_{w2}, X_{wm}, Y_{wm} \)
    
\end{algorithmic}
\end{algorithm}

\subsubsection{\textbf{DW Watermark Insertion}}
At the end of legalization with placement $P_{lg} = (X_{lg}, Y_{lg})$, cells are moved along the x- or y-axis to insert unique watermark signatures hashed as $N$ bit sequences $B_{N}$. If the $i$-th bit in the signature $B_{N}$ is 1, the cell $c_{i}$ randomly selected from the candidate watermarked cell set $C_x$ is moved along the x-axis for $D_x[c_i]$. If the $i$-th signature bit is 0, the cell $c_{i}$ is randomly chosen from the candidate watermarked cell set $C_y$ and moved along the y-axis for $D_y[c_i]$. 
The resulting placement is denoted as intermediate placement $P_{itr}=(X_{itr}, Y_{itr})$, and the indices of the cells moved along the x- or y-axis form the watermarked cells $C_{w2}$. 
These movements are performed before detailed placement, allowing possible performance degradation to be compensated during the subsequent detailed placement phase and yield a watermarked solution $P_{wm}=(X_{wm}, Y_{wm})$.
The selected watermarked cells $C_{w2}$, and their corresponding position distance $\mathrm{Dist} = P_{itr}(C_{w2})-P_{wm}(C_{w2})$ between intermediate placement $P_{itr}(C_{w2})$ and the watermarked placement $P_{wm}(C_{w2})$ constitute the watermark. \revision{The random wm cell selection before detailed placement ensures the signatures are encoded randomly.}

\subsubsection{\textbf{DW Watermark Extraction}}
The design owner detects watermarks in the placement $P^{\prime}$ by comparing the watermarked cell $C_{w2}$ position with the intermediate placement $P_{itr}$ and calculates the new distance $\mathrm{Dist}^{\prime}$ as $P_{itr}(C_{w2})-P^{\prime}(C_{w2})$. If $C_{w2}^\prime$ cells in the extracted $\mathrm{Dist}^{\prime}$ matches the $\mathrm{Dist}$ both along x- and y-axis, the watermark extraction rate is calculated as Eqn.~\ref{eq:dw_wer}.

\begin{equation}
    \label{eq:dw_wer}
   \%WER_{DW} = 100\times\frac{|C_{w2}^\prime|}{|C_{w2}|}
\end{equation}
 
\subsection{\sys{} Watermarking} \label{sec:sys}
As a combination of both the GW and the DW, \sys{} applies global watermarking during its global placement stage and detailed watermarking on top of the watermarked region before its detailed placement stage. By combining these two watermarking schemes, the inserted watermark is further augmented. 

\begin{figure}[ht]
    \centering
     
    \includegraphics[width= 0.95\columnwidth]{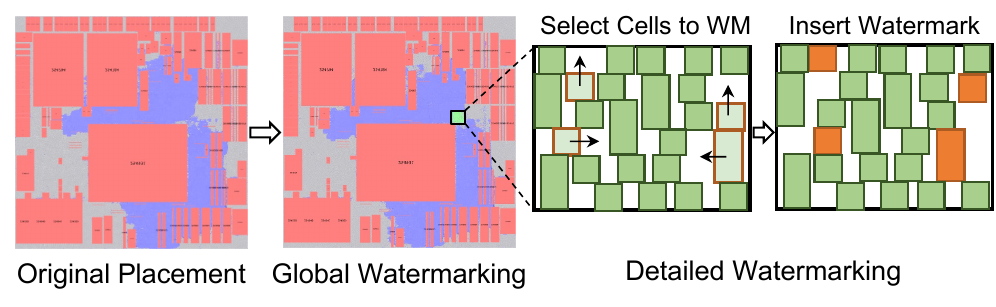}
 
     \caption{\sys{} framework. \sys{} first applies global watermarking during its global placement and then applies detailed watermarking on top of the watermarked region before the detailed placement. The \textcolor{green}{green} cells are the wm cells in GW stage and \textcolor{orange}{orange} cells are the wm cells in DW stage.}
   
    \label{fig:unimarks}
\end{figure}
\subsubsection{\textbf{\sys{} Watermark Insertion}}
As in Fig.~\ref{fig:unimarks}, \sys{} employs a hybrid approach that combines \textit{Global} and \textit{Detailed Watermarking} to produce stronger watermarks. Initially, \sys{} searches for a watermarked region using the strategy akin to \textit{Global Watermarking}, yielding a watermarked region $R_w$ and associated cells $C_{w1}$. \sys{} runs global placement and legalization with such watermark constraints. Then, it perturbs cells along the x- and y-axis within the watermarked region $R_w$, following the approach of \textit{Detailed Watermarking} and obtain an intermediate placement $P_{itr}$. Afterward, the detailed placement proceeds to compensate for watermarking degradation and produce the watermarked placement $P_{wm}$. The perturbed distance $\mathrm{Dist}$ and cells $C_{w2}$ are the watermarks in the second step. 

\subsubsection{\textbf{\sys{} Watermark Extraction}}
To claim ownership of the layout, given the cell positions $P^\prime$, the watermark holder can use the following two steps. (i) The extraction of the watermarks $C_{w1}^{\prime}$ from the watermarked region $R_w$, with the global extraction rate denoted as in Eqn.~\ref{eq:gw_wer}. (ii) Afterwards, the inserted signatures are extracted by comparing $P^\prime - P_{itr}$ with $\mathrm{Dist}$ at the watermarked cell indices $C_{w2}$. The number of cells successfully matched is denoted as $C_{w2}^\prime$, from which the detailed extraction rate is calculated as in Eqn.~\ref{eq:dw_wer}.  
The final WER is calculated in Eqn.~\ref{eq:unimark_wer}.

\begin{equation}
\label{eq:unimark_wer}
\begin{array}{ll}
\%WER_{\sys} = 100 \times \frac{WER_{GW} + WER_{DW}}{2}
\end{array}
\end{equation}

\section{Experiments}
\label{sec:exp}
In this section, we first introduce the experiment setups in Section~\ref{subsec:setup}. Then, we evaluate the performance of wirelength-driven and timing-driven design watermarking in Section~\ref{subsec:results}. Next, we show the robustness of \sys{} under attacks in Section~\ref{sec:security_ana}.  
\subsection{Experimental Setup}~\label{subsec:setup}
\textbf{Datasets}:
We evaluate the watermarking impact on wirelength-driven benchmark suites: ISPD'2015~\cite{bustany2015ispd} and ISPD'2019~\cite{liu2019ispd}; and the timing-driven ICCAD'2015~\cite{7372671} benchmark suite. All designs in the ICCAD'2015 benchmark suite are large and contain more than 700k cells. 
The statistics and \revision{node technology} of these designs are summarized in Table~\ref{tab:dataset_statistic}, with designs having fence regions marked in {\color[HTML]{4472C4} blue}. The macros, fence regions, and IO pin locations in the benchmarks are fixed.


\begin{table}[!ht]
\resizebox{\columnwidth}{!}{%
\begin{tabular}{|c|c|c|c|c|c|c|}
\hline
Suite & Design & Cells & Nets & Size & Stride &  Node\\ \hline
\hline 
 & {\color[HTML]{4472C4} perf\_a} & 108K & 115K & 10 & 5 & 65nm\\ \cline{2-7} 
 & {\color[HTML]{4472C4} perf\_b} & 113K & 113K & 10 & 5  & 65nm\\ \cline{2-7} 
 & {\color[HTML]{4472C4} dist\_a} & 127K & 134K & 20 & 5& 65nm\\ \cline{2-7} 
 & {\color[HTML]{4472C4} mult\_b} & 146K & 152K & 10 & 5 & 65nm\\ \cline{2-7} 
 & {\color[HTML]{4472C4} mult\_c} & 146K & 152K & 20 & 5  & 65nm\\ \cline{2-7} 
 & {\color[HTML]{4472C4} pci\_a} & 30K & 34K & 10 & 5 & 65nm \\ \cline{2-7} 
 & {\color[HTML]{4472C4} pci\_b} & 29K & 33K & 20 & 5& 65nm \\ \cline{2-7} 
 & {\color[HTML]{4472C4} superblue11} & 926K & 936K & 100 & 40  & 28nm\\ \cline{2-7} 
 & {\color[HTML]{4472C4} superblue16} & 680K & 697K & 50 & 10 & 28nm\\ \cline{2-7} 
 & perf\_1 & 113K & 113K & 50 & 10 & 65nm\\ \cline{2-7} 
 & fft\_1 & 35K & 33K & 10 & 5 & 65nm\\ \cline{2-7} 
 & fft\_2 & 35K & 33K & 10 & 5 & 65nm \\ \cline{2-7} 
 & fft\_a & 34K & 32K & 10 & 5 & 65nm\\ \cline{2-7} 
 & fft\_b & 34K & 32K & 20 & 5 & 65nm \\ \cline{2-7} 
 & mult\_1 & 160K & 159K & 50 & 10 & 65nm \\ \cline{2-7} 
 & mult\_2 & 160K & 159K & 50 & 10 & 65nm\\ \cline{2-7} 
 & mult\_a & 154K & 154K & 50 & 10& 65nm\\ \cline{2-7} 
 & superblue12 & 1293K & 1293K & 50 & 10 & 28nm \\ \cline{2-7} 
 & superblue14 & 634K & 620K & 50 & 10 & 28nm \\ \cline{2-7} 
\multirow{-20}{*}{\begin{tabular}[c]{@{}c@{}}ISPD\\ 2015\\ (WL)\end{tabular}} & superblue19 & 522K & 512K & 100 & 10& 28nm  \\ \hline
\hline 
 & ispd19test1 & 9K & 3K & 10 & 5 & 32nm \\ \cline{2-7} 
 & ispd19test2 & 73K & 72K & 10 & 5& 32nm  \\ \cline{2-7} 
 & ispd19test3 & 8K & 9K & 10 & 1 & 32nm \\ \cline{2-7} 
 & ispd19test4 & 151K & 146K & 10 & 5& 65nm \\ \cline{2-7} 
 & {\color[HTML]{4472C4} ispd19test5} & 29K & 29K & 10 & 10& 65nm  \\ \cline{2-7} 
 & ispd19test6 & 181K & 180K & 50 & 10 & 32nm \\ \cline{2-7} 
 & ispd19test7 & 362K & 359K & 50 & 25& 32nm  \\ \cline{2-7} 
 & ispd19test8 & 543K & 538K & 50 & 10& 32nm \\ \cline{2-7} 
 & ispd19test9 & 903K & 895K & 200 & 20& 32nm \\ \cline{2-7} 
\multirow{-10}{*}{\begin{tabular}[c]{@{}c@{}}ISPD\\ 2019\\ (WL)\end{tabular}} & ispd19test10 & 903K & 895K & 10 & 5 & 32nm\\ \hline
\hline 
 & superblue1 & 1209k & 1215k & 100 & 50&45nm \\ \cline{2-7} 
 & superblue3 & 1213k & 1224k & 100 & 30&45nm  \\ \cline{2-7} 
 & superblue4 & 795k & 802k & 100 & 50 &45nm\\ \cline{2-7} 
 & superblue5 & 1086k & 1100k & 100 & 50&45nm \\ \cline{2-7} 
 & superblue7 & 1931k & 1933k & 100 & 30&45nm  \\ \cline{2-7} 
 & superblue10 & 1876k & 1898k & 100 & 30&45nm \\ \cline{2-7} 
 & superblue16 & 981k & 999k & 100 & 30&45nm  \\ \cline{2-7} 
\multirow{-8}{*}{\begin{tabular}[c]{@{}c@{}}ICCAD\\ 2015\\ (Timing)\end{tabular}} & superblue18 & 768k & 771k & 100 & 50&45nm \\ \hline
\end{tabular}%
}
\caption{Benchmark statistics and node technology. The designs with fence regions are in {\color[HTML]{4472C4} blue}. The sliding window's size and stride are multiples of the adjacent row-height. }

\label{tab:dataset_statistic}
\end{table}

\textbf{Experiment Setup}: \sys{} is implemented with Python 3.9 and benchmarked on a Linux Ubuntu system equipped with NVIDIA TITAN XP GPUs, each with 12 GB RAM, and 48 Intel(R) Xeon(R) CPUs, accompanied by 128 GB RAM.
The watermarking methodology of \sys{} can be integrated into any VLSI physical design framework, and we chose to build upon the open-source placement framework, DREAMPlace~\cite{gu2020dreamplace} to demonstrate its viability.
The DREAMPlace~\cite{gu2020dreamplace} accelerates the state-of-the-art placement algorithm ePlace~\cite{lu2015eplace}/RePlAce~\cite{cheng2018replace} on a GPU and maintains the same level of performance. 
For routing, we employ the open-source state-of-the-art framework CUGR~\cite{9218646}.
For timing-driven benchmark suite, we use DREAMPlace 4.0~\cite{9774725} where the timing metrics are measured by OpenTimer~\cite{huang2020opentimer}.  

\textbf{Hyperparameters}:
For the hyperparameters, we perform a grid search for the GW phase. In the search of Eqn.~\ref{eq:score}, the $\alpha$ = \{0.1, 0.5, 0\}, $\beta$=\{0.1, 0.5, 0\}, and $\gamma$ = 0.1. The search terminates when the watermarked layout quality is not compromised.
For all the considered benchmark suites, we report the GW hyperparameters in 
Table~\ref{tab:dataset_statistic}.
In the DW phase, $d_x = 1$, and $d_y$ is set to one adjacent row-height.

\textbf{Baseline}: We choose the following state-of-the-art watermarking frameworks as our baselines: (i) topology constraint-based \textbf{Row Parity}~\cite{kahng1998robust,kahng2001constraint} that inserts unique bit sequences as watermarks by shifting cells to different rows in the placement stage. Cells with a 1-bit are moved to an odd row, while cells with a 0-bit are moved to an even row; (ii) position constraint-based \textbf{Cell Scattering}~\cite{4415879} that employs pseudorandom coordinate transformation (PRCT) algorithms to scatter the watermarked cells on the chip canvas as watermarks. Cells with 1-bit are moved along the y-axis, and cells with 0-bit are moved along the x-axis if they do not overlap with their neighbors; (iii) invasive \textbf{Buffer Insertion}~\cite{sun2006watermarking} that adds additional buffers as watermarks without affecting timing critical paths during the placement stage. Two buffers are inserted to represent 0-bit, and one buffer is inserted to represent 1-bit. \revision{For fair comparisons, 
we re-implemented the baselines~\cite{kahng1998robust,kahng2001constraint,4415879,sun2006watermarking} and integrated them into the DREAMPlace codebase for benchmarking.}

We skip the baselines that: (i) have different watermarking targets. ~\cite{ni2004watermarking,bai2007watermarking} are designed for smaller full-custom IC designs, ~\cite{saha2007fast,liang2011chaotic} are designed for FPGAs, whereas our watermarking targets are modern digital VLSI designs; (ii) have similar watermarking approaches as our baselines, and we use the baselines as a proof-of-concept. ~\cite{saha2007novel} inserts flip-flops instead of buffers into the layouts as watermarks. 
The signature length is set to 50-bit for all frameworks.

\begin{table*}[!ht]
\centering
\resizebox{0.97\textwidth}{!}{%
\begin{tabular}{|c|c|cc|cc|cc|cc|cc|cc|}
\hline
& & \multicolumn{2}{c|}{Row Parity~\cite{kahng1998robust,kahng2001constraint}} & \multicolumn{2}{c|}{\baseconstcap~\cite{4415879}}& \multicolumn{2}{c|}{\baseinvasivecap~\cite{sun2006watermarking}} & \multicolumn{2}{c|}{\sys: GW} & \multicolumn{2}{c|}{\sys: DW} & \multicolumn{2}{c|}{\sys{}: GW+DW} \\ \cline{3-14} 
\multirow{-2}{*}{Design} & \multirow{-2}{*}{Utl.} & \multicolumn{1}{c|}{PWLR $\downarrow$} & RWLR $\downarrow$ & \multicolumn{1}{c|}{PWLR $\downarrow$} & RWLR $\downarrow$ & \multicolumn{1}{c|}{PWLR $\downarrow$} & RWLR $\downarrow$ & \multicolumn{1}{c|}{PWLR $\downarrow$} & RWLR $\downarrow$ & \multicolumn{1}{c|}{PWLR $\downarrow$} & RWLR $\downarrow$ & \multicolumn{1}{c|}{PWLR $\downarrow$} & RWLR $\downarrow$ \\ \hline
\hline 
{\color[HTML]{4472C4} perf\_a $\star$} & 71.69\% & \multicolumn{1}{c|}{1.0045} & 1.0155 &\multicolumn{1}{c|}{0.9978} & 0.9967 & \multicolumn{1}{c|}{\textcolor{gray}{1.5289}} &  \textcolor{gray}{2.3270} & \multicolumn{1}{c|}{0.9976} & 0.9873 & \multicolumn{1}{c|}{0.9994} & 1.0003 & \multicolumn{1}{c|}{\textbf{0.9972}} & \textbf{0.9873} \\ \hline
{\color[HTML]{4472C4} perf\_b} & 49.71\% & \multicolumn{1}{c|}{1.0058} & 1.0267&  \multicolumn{1}{c|}{1.0020} & 1.0001 & \multicolumn{1}{c|}{1.0176} & 1.0745 & \multicolumn{1}{c|}{0.9930} & 0.9901 & \multicolumn{1}{c|}{0.9990} & 1.0002 & \multicolumn{1}{c|}{\textbf{0.9890}} & \textbf{0.9901} \\ \hline
{\color[HTML]{4472C4} dist\_a} & 61.62\% & \multicolumn{1}{c|}{1.0015} & 0.9999 & \multicolumn{1}{c|}{0.9984} & 0.9965 & \multicolumn{1}{c|}{1.0995} & 1.1072 & \multicolumn{1}{c|}{1.0004} & 1.0052 & \multicolumn{1}{c|}{0.9998} & 1.0005 & \multicolumn{1}{c|}{\textbf{1.0004}} & \textbf{1.0052} \\ \hline
{\color[HTML]{4472C4} mult\_b $\star$} & 77.31\% & \multicolumn{1}{c|}{1.0047} & 1.0199 & \multicolumn{1}{c|}{0.9991}  & 0.9923& \multicolumn{1}{c|}{\textcolor{gray}{1.8966}} & \textcolor{gray}{2.9374}  & \multicolumn{1}{c|}{1.0010} & 1.0028 & \multicolumn{1}{c|}{1.0002} & 1.0044 & \multicolumn{1}{c|}{\textbf{0.9994}} & \textbf{1.0028} \\ \hline
{\color[HTML]{4472C4} mult\_c $\star$} & 77.31\% & \multicolumn{1}{c|}{1.0031} & 1.0252 & \multicolumn{1}{c|}{0.9980}& 
1.0023& \multicolumn{1}{c|}{\textcolor{gray}{1.6003}} & \textcolor{gray}{2.7459}  &  \multicolumn{1}{c|}{1.0017} & 0.9979 & \multicolumn{1}{c|}{0.9990} & 1.0020 & \multicolumn{1}{c|}{\textbf{0.9963}} & \textbf{0.9979} \\ \hline
{\color[HTML]{4472C4} pci\_a} & 50.84\% & \multicolumn{1}{c|}{1.0080} & 1.0340 & \multicolumn{1}{c|}{0.9931} & 0.9846 & \multicolumn{1}{c|}{\textcolor{gray}{1.6269}} & \textcolor{gray}{1.6849}& \multicolumn{1}{c|}{1.0361} & 1.0048 & \multicolumn{1}{c|}{1.0011} & 1.0021 & \multicolumn{1}{c|}{\textbf{0.9997}} & \textbf{0.9950} \\ \hline
{\color[HTML]{4472C4} pci\_b }& 54.77\% & \multicolumn{1}{c|}{1.0069} & 1.1173 &\multicolumn{1}{c|}{0.9912}  & 0.9977 & \multicolumn{1}{c|}{\textcolor{gray}{1.2239}} & \textcolor{gray}{1.8207} & \multicolumn{1}{c|}{0.9951} & 1.0026 & \multicolumn{1}{c|}{0.9997} & 1.0008 & \multicolumn{1}{c|}{\textbf{0.9951}} & \textbf{1.0026} \\ \hline
{\color[HTML]{4472C4} superblue11 $\star$} & 73.64\% & \multicolumn{1}{c|}{1.0052} & 1.0344 &  \multicolumn{1}{c|}{1.0278} & 1.0358 & \multicolumn{1}{c|}{\textcolor{gray}{1.6930}} & \textcolor{gray}{3.7086} & \multicolumn{1}{c|}{0.9992} & 0.9992 & \multicolumn{1}{c|}{0.9995} & 1.0005 & \multicolumn{1}{c|}{\textbf{0.9986}} & \textbf{0.9992} \\ \hline
{\color[HTML]{4472C4} superblue16 $\star$} &74.49\% & \multicolumn{1}{c|}{1.0030} & 1.0210 & \multicolumn{1}{c|}{0.9988} & 0.9989 & \multicolumn{1}{c|}{1.1023} & 1.1377 & \multicolumn{1}{c|}{1.0018} & 1.0204 & \multicolumn{1}{c|}{\textbf{1.0000}} & \textbf{1.0000} & \multicolumn{1}{c|}{1.0017} & 1.0204 \\ \hline
perf\_1 $\star$ & 90.57\% & \multicolumn{1}{c|}{1.0035} & 1.0199 & \multicolumn{1}{c|}{0.9985} & 0.9983 & \multicolumn{1}{c|}{1.0150} &  1.0642 &  \multicolumn{1}{c|}{1.0031} & 0.9973 & \multicolumn{1}{c|}{1.0042} & 1.0019 & \multicolumn{1}{c|}{\textbf{0.9964}} & \textbf{0.9973} \\ \hline
fft\_1 $\star$ & 83.54\% & \multicolumn{1}{c|}{1.0017} & 1.0260 & \multicolumn{1}{c|}{1.0167} & 1.0156 & \multicolumn{1}{c|}{1.0680} &1.1457 & \multicolumn{1}{c|}{0.9671} & 0.9673 & \multicolumn{1}{c|}{1.0026} & 0.9975 & \multicolumn{1}{c|}{\textbf{0.9671}} & \textbf{0.9673} \\ \hline
fft\_2  & 49.97\% & \multicolumn{1}{c|}{1.0050} & 1.0260 & \multicolumn{1}{c|}{1.0148}& 1.0114 & \multicolumn{1}{c|}{\textcolor{gray}{1.4603}} & \textcolor{gray}{1.4476}  &  \multicolumn{1}{c|}{0.9767} & 0.9770 & \multicolumn{1}{c|}{1.0027} & 0.9996 & \multicolumn{1}{c|}{\textbf{0.9767}} & \textbf{0.9770} \\ \hline
fft\_a $\star$ & 74.02\% & \multicolumn{1}{c|}{1.0121} & 1.0200 & \multicolumn{1}{c|}{1.0024}& 0.9933& \multicolumn{1}{c|}{\textcolor{gray}{1.8203}} & \textcolor{gray}{2.1923}  &  \multicolumn{1}{c|}{0.9939} & 0.9895 & \multicolumn{1}{c|}{1.0075} & 1.0094 & \multicolumn{1}{c|}{\textbf{0.9939}} & \textbf{0.9895} \\ \hline
fft\_b $\star$& 78.01\% & \multicolumn{1}{c|}{1.0018} & 1.0075 & \multicolumn{1}{c|}{0.9961} &1.0030 & \multicolumn{1}{c|}{\textcolor{gray}{1.9859}} & \textcolor{gray}{2.5615} &  \multicolumn{1}{c|}{0.9909} & 0.9890 & \multicolumn{1}{c|}{0.9991} & 1.0036 & \multicolumn{1}{c|}{\textbf{0.9909}} & \textbf{0.9890} \\ \hline
mult\_1 $\star$ & 80.24\%& \multicolumn{1}{c|}{1.0050} & 1.0238 & \multicolumn{1}{c|}{1.0077} & 1.0065 & \multicolumn{1}{c|}{1.0464} &1.0631  &  \multicolumn{1}{c|}{0.9753} & 0.9744 & \multicolumn{1}{c|}{1.0004} & 1.0008 & \multicolumn{1}{c|}{\textbf{0.9753}} & \textbf{0.9744} \\ \hline
mult\_2 $\star$& 79.03\% & \multicolumn{1}{c|}{1.0033} & 1.0199 &  \multicolumn{1}{c|}{0.9984} & 0.9967 & \multicolumn{1}{c|}{1.0736} & 1.1147&  \multicolumn{1}{c|}{0.9867} & 0.9900 & \multicolumn{1}{c|}{0.9990} & 0.9992 & \multicolumn{1}{c|}{\textbf{0.9852}} & \textbf{0.9900} \\ \hline
mult\_a $\star$& 86.10\% &\multicolumn{1}{c|}{1.0037} & 1.0105 & \multicolumn{1}{c|}{ 0.9995}& 0.9968 & \multicolumn{1}{c|}{\textcolor{gray}{1.3862}} & \textcolor{gray}{1.6738}  &  \multicolumn{1}{c|}{0.9973} & 0.9916 & \multicolumn{1}{c|}{1.0005} & 0.9958 & \multicolumn{1}{c|}{\textbf{0.9973}} & \textbf{0.9916} \\ \hline
superblue12 & 57.62\% & \multicolumn{1}{c|}{1.0031} & 1.0067 &\multicolumn{1}{c|}{0.9979} & 0.9956 & \multicolumn{1}{c|}{1.0683} & 1.1044 &  \multicolumn{1}{c|}{1.0036} & 0.9732 & \multicolumn{1}{c|}{0.9994} & 0.9942 & \multicolumn{1}{c|}{\textbf{0.9854}} & \textbf{0.9732} \\ \hline
superblue14 $\star$& 77.63\%& \multicolumn{1}{c|}{1.0020} & 1.0057 & \multicolumn{1}{c|}{0.9991} & 0.9981 & \multicolumn{1}{c|}{1.0212} & 1.0286 & 
 \multicolumn{1}{c|}{\textbf{0.9851}} & 0.9867 & \multicolumn{1}{c|}{1.0001} & 1.0000 & \multicolumn{1}{c|}{0.9887} & \textbf{0.9867} \\ \hline
superblue19 $\star$& 81.51\% &\multicolumn{1}{c|}{1.0025} & 1.0077 & \multicolumn{1}{c|}{1.0001} & 1.0005& \multicolumn{1}{c|}{1.0295} & 1.0672&  \multicolumn{1}{c|}{0.9881} & 0.9809 & \multicolumn{1}{c|}{1.0003} & 1.0001 & \multicolumn{1}{c|}{\textbf{0.9814}} & \textbf{0.9809} \\ \hline
\hline 
{\color[HTML]{4472C4} Average: FR} &- & \multicolumn{1}{c|}{1.0047} & 1.0322 & \multicolumn{1}{c|}{1.0006} & 1.0005 & \multicolumn{1}{c|}{1.0724} & 1.1061&  \multicolumn{1}{c|}{1.0028} & 1.0011 & \multicolumn{1}{c|}{0.9997} & 1.0012 & \multicolumn{1}{c|}{\textbf{0.9975}} & \textbf{1.0000} \\ \hline
Average: All&- &  \multicolumn{1}{c|}{1.0043} & 1.0231 & \multicolumn{1}{c|}{1.0018} & 1.0012 & \multicolumn{1}{c|}{1.0536} & 1.0901&  \multicolumn{1}{c|}{0.9946} & 0.9913 & \multicolumn{1}{c|}{1.0007} & 1.0006 & \multicolumn{1}{c|}{\textbf{0.9908}} & \textbf{0.9908} \\ \hline
\end{tabular}%
}
\caption{Performance on the ISPD'2015 benchmarks~\cite{bustany2015ispd}. All the design watermarks are successfully extracted, i.e., WER = 100\%. The PWLR and RWLR are the placement and routed wirelength rates over the original designs. FR is fence region. The results in \textcolor{gray}{gray} fail buffer insertion WM with significant degradation on the high-utilized designs (denoted with $\star$).
}
\label{tab:ispd2015}
\end{table*}

\begin{table*}[!ht]
\centering
\resizebox{0.97\textwidth}{!}{%
\begin{tabular}{|c|c|cc|cc|cc|cc|cc|cc|}
\hline
& & \multicolumn{2}{c|}{Row Parity~\cite{kahng1998robust,kahng2001constraint}} & \multicolumn{2}{c|}{\baseconstcap~\cite{4415879}}& \multicolumn{2}{c|}{\baseinvasivecap~\cite{sun2006watermarking}} & \multicolumn{2}{c|}{\sys: GW} & \multicolumn{2}{c|}{\sys: DW} & \multicolumn{2}{c|}{\sys{}: GW+DW} \\ \cline{3-14} 
\multirow{-2}{*}{Design} & \multirow{-2}{*}{Utl.} & \multicolumn{1}{c|}{PWLR $\downarrow$} & RWLR $\downarrow$ & \multicolumn{1}{c|}{PWLR $\downarrow$} & RWLR $\downarrow$ & \multicolumn{1}{c|}{PWLR $\downarrow$} & RWLR $\downarrow$ & \multicolumn{1}{c|}{PWLR $\downarrow$} & RWLR $\downarrow$ & \multicolumn{1}{c|}{PWLR $\downarrow$} & RWLR $\downarrow$ & \multicolumn{1}{c|}{PWLR $\downarrow$} & RWLR $\downarrow$ \\ \hline
\hline 
ispd19test1 $\star$& 83.14\% & \multicolumn{1}{c|}{1.0060} & 1.0129 & \multicolumn{1}{c|}{0.9978} & 0.9998 & \multicolumn{1}{c|}{1.0394} & 1.0619 & \multicolumn{1}{c|}{0.9995} & 1.0015 & \multicolumn{1}{c|}{1.0023} & 1.0043 & \multicolumn{1}{c|}{\textbf{0.9955}} & \textbf{1.0015} \\ \hline
ispd19test2 $\star$& 84.11\%& \multicolumn{1}{c|}{1.0184} & 1.0201 & \multicolumn{1}{c|}{1.0096} & 1.0016 & \multicolumn{1}{c|}{1.0127} & 1.0408 & \multicolumn{1}{c|}{\textbf{0.9981}} & 0.9999 & \multicolumn{1}{c|}{1.0021} & 1.0003 & \multicolumn{1}{c|}{0.9988} & \textbf{0.9999} \\ \hline
ispd19test3 $\star$& 88.77\% & \multicolumn{1}{c|}{1.0130} & 1.0475 & \multicolumn{1}{c|}{1.0180} & 1.0193 & \multicolumn{1}{c|}{1.0582} & 1.0801 & \multicolumn{1}{c|}{1.0059} & 1.0045 & \multicolumn{1}{c|}{\textbf{0.9961}} & \textbf{0.9976} & \multicolumn{1}{c|}{1.0059} & 1.0045 \\ \hline
ispd19test4 & 65.90\% &\multicolumn{1}{c|}{1.0017} & 1.0050 & \multicolumn{1}{c|}{0.9999} & 0.9998 & \multicolumn{1}{c|}{1.1452} & 1.2494 & \multicolumn{1}{c|}{0.9957} & 0.9907 & \multicolumn{1}{c|}{0.9998} & 1.0028 & \multicolumn{1}{c|}{\textbf{0.9957}} & \textbf{0.9907} \\ \hline
{\color[HTML]{4472C4} ispd19test5} & 43.06\% & \multicolumn{1}{c|}{1.0102} & 1.0556 & \multicolumn{1}{c|}{1.0998} & 1.0975 & \multicolumn{1}{c|}{0.9859}  & 1.0121 & \multicolumn{1}{c|}{1.0013} & 0.9998 & \multicolumn{1}{c|}{\textbf{0.9996}} & \textbf{0.9968} & \multicolumn{1}{c|}{1.0013} & 0.9998 \\ \hline
ispd19test6 $\star$& 74.60\% & \multicolumn{1}{c|}{1.0032} & 1.0106 & \multicolumn{1}{c|}{0.9994} & 0.9993 & \multicolumn{1}{c|}{1.0044} & 1.1108 & \multicolumn{1}{c|}{1.0023} & 1.0026 & \multicolumn{1}{c|}{\textbf{1.0001}} & \textbf{0.9997} & \multicolumn{1}{c|}{1.0023} & 1.0026 \\ \hline
ispd19test7 $\star$& 95.56\% &\multicolumn{1}{c|}{1.0028} & 1.0160 & \multicolumn{1}{c|}{1.0136} & 1.0109 & \multicolumn{1}{c|}{1.0003} & 1.0717 & \multicolumn{1}{c|}{1.0050} & 1.0054 & \multicolumn{1}{c|}{\textbf{1.0000}} & \textbf{0.9997} & \multicolumn{1}{c|}{1.0050} & 1.0054 \\ \hline
ispd19test8 $\star$& 79.90\% & \multicolumn{1}{c|}{1.0001} & 1.0082 &  \multicolumn{1}{c|}{0.9966} & 0.9958 & \multicolumn{1}{c|}{1.0118} & 1.0806 & \multicolumn{1}{c|}{0.9961} & 0.9929 & \multicolumn{1}{c|}{0.9999} & 0.9996 & \multicolumn{1}{c|}{\textbf{0.9961}} & \textbf{0.9929} \\ \hline
ispd19test9 $\star$& 84.02\% & \multicolumn{1}{c|}{1.0072} & 1.0108 & \multicolumn{1}{c|}{1.0108} & 1.0095 & \multicolumn{1}{c|}{1.0164} & 1.0814 & \multicolumn{1}{c|}{1.0023} & 1.0023 & \multicolumn{1}{c|}{\textbf{1.0009}} & \textbf{0.9999} & \multicolumn{1}{c|}{1.0023} & 1.0023 \\ \hline
ispd19test10 $\star$& 88.48\% & \multicolumn{1}{c|}{1.0025} & 1.0090 & \multicolumn{1}{c|}{1.0043} & 1.0108 & \multicolumn{1}{c|}{1.0378} & 1.0982 & \multicolumn{1}{c|}{0.9972} & 0.9967 & \multicolumn{1}{c|}{0.9999} & 1.0002 & \multicolumn{1}{c|}{\textbf{0.9972}} & \textbf{0.9966} \\ \hline
\hline 
Average & - & \multicolumn{1}{c|}{1.0065} & 1.0194 & \multicolumn{1}{c|}{1.0060}& 1.0140 & \multicolumn{1}{c|}{1.0304} &1.0871 & \multicolumn{1}{c|}{1.0003} & 0.9996 & \multicolumn{1}{c|}{1.0001} & 1.0001 & \multicolumn{1}{c|}{\textbf{1.0000}} & \textbf{0.9996} \\ \hline
\end{tabular}%
}
\caption{Performance on ISPD'2019 benchmarks~\cite{liu2019ispd}. All the design watermarks are successfully extracted, i.e., WER = 100\%. The PWLR and RWLR are the placement and routed wirelength rates over the original designs.}
\vspace{-5pt}
\label{tab:ispd2019}
\end{table*}




\textbf{Evaluation Metrics}:
For the wirelength-driven benchmark suites, whose quality is measured by the layout wirelength, we use three metrics to evaluate the watermark performance:
(i) \textbf{Placement WireLength Rate (PWLR)}: The rate of estimated half-perimeter wirelength (HPWL) of watermarked layout compared to the original one; (ii) \textbf{Routing WireLength Rate (RWLR)}: The rate of routed wirelength of watermarked layout compared to the original one; (iii) \textbf{Watermark Extraction Rate (WER)}: The percentage of signatures extracted from the watermarked layout.
For the timing-driven benchmark suite, whose quality is measured by the timing slack, we use three metrics to evaluate the performance under static timing analysis~\cite{huang2020opentimer}: (i) \textbf{Total Negative Slack Rate (TNSR)}: The rate of total negative slack (TNS) of watermarked layout compared to original one; (ii) \textbf{Worst Negative Slack Rate (WNSR)}:The rate of worst negative slack (WNS) of watermarked layout compared to original one; (iii) \textbf{Watermark Extraction Rate (WER)}: The percentage of signatures extracted from the layout. The rate average is calculated by the geometric mean of designs' metrics. 

 \begin{table*}[!ht]
\centering
\resizebox{0.97\textwidth}{!}{%
\begin{tabular}{|c|c|cc|cc|cc|cc|cc|cc|}
\hline
& & \multicolumn{2}{c|}{Row Parity~\cite{kahng1998robust,kahng2001constraint}} & \multicolumn{2}{c|}{\baseconstcap~\cite{4415879}}& \multicolumn{2}{c|}{\baseinvasivecap~\cite{sun2006watermarking}} & \multicolumn{2}{c|}{\sys: GW} & \multicolumn{2}{c|}{\sys: DW} & \multicolumn{2}{c|}{\sys: GW+DW} \\ \cline{3-14} 
\multirow{-2}{*}{Design} & \multirow{-2}{*}{Utl.} &  \multicolumn{1}{c|}{TNSR $\downarrow$} & WNSR $\downarrow$ & \multicolumn{1}{c|}{TNSR $\downarrow$} & WNSR $\downarrow$ & \multicolumn{1}{c|}{TNSR $\downarrow$} & WNSR $\downarrow$ & \multicolumn{1}{c|}{TNSR $\downarrow$} & WNSR $\downarrow$ & \multicolumn{1}{c|}{TNSR $\downarrow$} & WNSR $\downarrow$ & \multicolumn{1}{c|}{TNSR $\downarrow$} & WNSR $\downarrow$ \\ \hline
\hline 
superblue1 & 78.07\% & \multicolumn{1}{c|}{1.0053} & \multicolumn{1}{c|}{1.0094} & \multicolumn{1}{c|}{1.0119} & 1.0026 & \multicolumn{1}{c|}{1.0083} & 1.0020 & \multicolumn{1}{c|}{0.9001} & \multicolumn{1}{c|}{\textbf{0.8514}}  & \multicolumn{1}{c|}{1.0256} & \multicolumn{1}{c|}{1.0249}  & \multicolumn{1}{c|}{\textbf{0.8283}} & \multicolumn{1}{c|}{1.0021} \\ \hline
superblue3 & 76.78\% & \multicolumn{1}{c|}{1.0053} & \multicolumn{1}{c|}{1.0176}  & \multicolumn{1}{c|}{1.0066} & 0.9891 & \multicolumn{1}{c|}{1.1035} & 1.0988 &  \multicolumn{1}{c|}{\textbf{0.9302}} & \multicolumn{1}{c|}{\textbf{0.9029}} & \multicolumn{1}{c|}{1.0035} & \multicolumn{1}{c|}{0.9884}  & \multicolumn{1}{c|}{0.9363} & \multicolumn{1}{c|}{0.9047} \\ \hline
superblue4 & 80.12\% & \multicolumn{1}{c|}{1.1299} & \multicolumn{1}{c|}{0.9914} & \multicolumn{1}{c|}{0.9698} & 1.0191 & \multicolumn{1}{c|}{1.1954} & 1.1995 & \multicolumn{1}{c|}{\textbf{0.9037}} & \multicolumn{1}{c|}{\textbf{0.9370}} & \multicolumn{1}{c|}{0.9554} & \multicolumn{1}{c|}{0.9930}  & \multicolumn{1}{c|}{0.9221} & \multicolumn{1}{c|}{1.0094}  \\ \hline
superblue5 & 73.30\%& \multicolumn{1}{c|}{0.9801} & \multicolumn{1}{c|}{\textbf{0.9887}} & \multicolumn{1}{c|}{1.0126} & 0.9977 & \multicolumn{1}{c|}{1.0932} &1.1174 &  \multicolumn{1}{c|}{0.9159} & \multicolumn{1}{c|}{1.0001} & \multicolumn{1}{c|}{0.9922} & \multicolumn{1}{c|}{0.9959}& \multicolumn{1}{c|}{\textbf{0.9159}} & \multicolumn{1}{c|}{1.0508} \\ \hline
superblue7 & 76.48\% & \multicolumn{1}{c|}{0.9801} & \multicolumn{1}{c|}{0.9932} & \multicolumn{1}{c|}{0.9979} &1.0212 & \multicolumn{1}{c|}{0.9933} & 1.0199 & \multicolumn{1}{c|}{\textbf{0.9584}} & \multicolumn{1}{c|}{0.9758}  & \multicolumn{1}{c|}{0.9934} & \multicolumn{1}{c|}{1.0199} & \multicolumn{1}{c|}{0.9636} & \multicolumn{1}{c|}{\textbf{0.9727}} \\ \hline
superblue10 & 75.80\% & \multicolumn{1}{c|}{1.0099} & \multicolumn{1}{c|}{\textbf{1.0041}} & \multicolumn{1}{c|}{1.0218} & 1.0141& \multicolumn{1}{c|}{1.0194} & 1.2918 & \multicolumn{1}{c|}{1.0069} & \multicolumn{1}{c|}{1.0072}   & \multicolumn{1}{c|}{\textbf{1.0058}} & \multicolumn{1}{c|}{1.0068}  & \multicolumn{1}{c|}{1.0069} & \multicolumn{1}{c|}{1.0072} \\ \hline
superblue16 & 79.23\% & \multicolumn{1}{c|}{0.9814} & \multicolumn{1}{c|}{1.1271} & \multicolumn{1}{c|}{0.9067} & 1.3274 & \multicolumn{1}{c|}{1.0109} & 1.0877 & \multicolumn{1}{c|}{1.0166} & \multicolumn{1}{c|}{1.0473} & \multicolumn{1}{c|}{\textbf{0.9060}} & \multicolumn{1}{c|}{1.3485}  & \multicolumn{1}{c|}{1.0069} & \multicolumn{1}{c|}{\textbf{1.0072}} \\ \hline
superblue18 & 67.44\% & \multicolumn{1}{c|}{1.0068} & \multicolumn{1}{c|}{0.9965}  & \multicolumn{1}{c|}{0.9981} & 1.0064  & \multicolumn{1}{c|}{1.0438} & 1.0613&  \multicolumn{1}{c|}{0.9337} & \multicolumn{1}{c|}{\textbf{0.9403}}  & \multicolumn{1}{c|}{1.0014} & \multicolumn{1}{c|}{1.0002}  & \multicolumn{1}{c|}{\textbf{0.9254}} & \multicolumn{1}{c|}{0.9449} \\ \hline
\hline
Average & - & \multicolumn{1}{c|}{1.0114} & \multicolumn{1}{c|}{1.0151} & \multicolumn{1}{c|}{0.9900} & 1.0424 & \multicolumn{1}{c|}{1.0566}& 1.1063 &\multicolumn{1}{c|}{0.9448} & \multicolumn{1}{c|}{\textbf{0.9559}}  & \multicolumn{1}{c|}{0.9848} & \multicolumn{1}{c|}{1.0418}   & \multicolumn{1}{c|}{\textbf{0.9366}} & \multicolumn{1}{c|}{0.9867} \\ \hline
\end{tabular}%
}
\caption{Performance on ICCAD'2015 benchmarks~\cite{7372671}. All the design watermarks are successfully extracted, i.e., WER = 100\%. The TNSR and WNSR are the total and worst negative slack rates over the original designs.}
\vspace{-10pt}
\label{tab:iccad2015}
\end{table*}


To preserve the optimized outcomes from the physical design algorithms, we set a threshold of layout quality degradation to not exceed 0.5\%. Because surpassing this threshold would counteract the benefits derived from the performance enhancement efforts in physical design~\cite{qiu2023progress}.

\subsection{Experimental Results}~\label{subsec:results}
\subsubsection{Wirelength-driven Watermarking Fidelity}
The performance of different watermarking schemes is tabulated in Table~\ref{tab:ispd2015} for the ISPD'2015 benchmarks~\cite{bustany2015ispd}, and in Table~\ref{tab:ispd2019} for the ISPD'2019 benchmarks~\cite{liu2019ispd}. 
For all the layouts in Table~\ref{tab:ispd2015}-Table~\ref{tab:ispd2019}, their inserted watermarks can be successfully extracted. Therefore, we evaluate how much layout performance is compromised to accommodate such watermark insertion among these watermarking frameworks.


\textbf{Comparison with Prior Constraint-based Watermarking}~\cite{kahng1998robust,kahng2001constraint,4415879}: Among the two benchmark suites, \sys{} results in no PWLR and RWLR degradation for accommodating the 50-bit signature. In contrast,
the topology constraint-based Row Parity~\cite{kahng1998robust,kahng2001constraint} changes the row index of the cells to encode watermarks but does not consider potential design constraints (e.g., fence regions and macros) in VLSI design. As a result, it results in 2.31\% and 1.94\% routed wirelength (RWLR) degradation on ISPD'2015~\cite{bustany2015ispd} and ISPD'2019~\cite{liu2019ispd}, respectively. The position constraint-based Cell Scattering~\cite{4415879} perturbs cell locations in the optimized layout for signature insertion. However, randomly selecting and moving watermarked cells after the optimization is done results in quality degradation, as reflected by the 0.12\% and 1.4\% RWLR degradation over the original design on ISPD'2015~\cite{bustany2015ispd} and ISPD'2019~\cite{liu2019ispd}.

\textbf{Comparison with Invasive Watermarking}~\cite{sun2006watermarking}: The invasive watermarking Buffer Insertion~\cite{sun2006watermarking} overlooks the additional design constraints, like fence regions and macros, in the modern VLSI design. The additional 50 buffers might be added close to such design constraints and result in significant cell displacement, as the design has to accommodate the buffers while satisfying the design constraints. As in Table~\ref{tab:ispd2015}-Table~\ref{tab:ispd2019}, Buffer Insertion~\cite{sun2006watermarking} introduced 9.01\% and 8.71\% routed wirelength (RWLR) degradation on ISPD'2015~\cite{bustany2015ispd} and ISPD'2019~\cite{liu2019ispd} benchmarks. \sys, however, has no wirelength degradation on both benchmarks. Besides, for the highly-utilized designs, where most of the layout space is filled with standard cell and fixed marcos/fence regions/IO pins, to accommodate the additional buffers, cells have to be displaced significantly from their non-watermarked position for buffer insertion. As such, it introduced significant PWLR and RWLR degradations as the grey numbers in Table~\ref{tab:ispd2015}. 

\textbf{DW Insertion Before/After Detailed Placement:}  After the placement stage is finished, the baseline Cell Scattering~\cite{4415879} moves cells in the layout along the x/y axis if there is space. The key difference is Cell Scattering~\cite{4415879} performs the movement after detailed placement, whereas the \sys: DW is performed before detailed placement. As seen in Table~\ref{tab:ispd2015}-~\ref{tab:ispd2019}, \sys: DW improves the RWLR quality from Cell Scattering~\cite{4415879} by 0.06\% and 1.39\%, respectively. 

\textbf{Comparison with the GW and DW submodules}: As a combination of GW and DW, \sys{} takes advantage of their strength and further reduces the performance degradation from accommodating watermarks. In contrast to GW, \sys{} slightly improves the placement quality. This improvement stems from modifying cell positions within a designated region, either across different rows or along the x-axis. Such modifications inherently disturb solutions formulated during global placement. Given that these disturbances occur in less compact regions, they provide different inputs to the detailed placement process, thereby offering potential optimization trajectories to minimize overall wirelength. 
The optimization to GW depends on the quality of the perturbations introduced, where the refinement on ISPD'2019~\cite{liu2019ispd} is marginal, and the improvement on  ISPD'2015~\cite{bustany2015ispd} is 0.05\% over GW.

Compared with DW, \sys{} watermarks upon the placement solution from GW, whereas DW watermarks the original solution. The additional co-optimization of the GW watermarked region makes the placer further refine the cell positions in the watermarked region and, thus, improves the watermarked layout quality. By encoding DW signature on the GW layout, \sys{} introduced no quality degradations. 
In contrast, DW watermarks upon the original placement solution, which could introduce more performance degradation when inserting DW signatures. It is reflected by the 0.06\% and 0.01\% RWLR improvement over the original design on ISPD'2015~\cite{bustany2015ispd} and ISPD'2019~\cite{liu2019ispd} benchmarks, respectively.

\revision{\textbf{Layout Utilization and Design Constraints Impact on WM Performance:} As in Table~\ref{tab:ispd2015}-~\ref{tab:iccad2015}, invasive watermarking yields worse performance on high layout utilization designs ($\geq$ 70\%), where the layout utilization is larger than 70\%. For invasive watermarking, the performance degradation becomes worse on designs with additional design constraints, e.g., fence regions and macros. Because the buffers can be encoded close to such constraints and result in significant cell displacements. In contrast, constraint-based watermarking approaches co-optimize the watermark and placement objectives to preserve the layout quality. }


 \subsubsection{Timing-driven Watermarking Fidelity}
As depicted in Table~\ref{tab:iccad2015}, \sys{} continuously preserves the layout quality in timing-driven watermarking benchmark ICCAD'2015~\cite{7372671}. By strategically searching for the watermarked region and cells with minimal impact on the performance, \sys{} results in no WNS and TNS degradation.
On the opposite, the constraint-based Row Parity~\cite{kahng1998robust,kahng2001constraint} and Cell Scattering~\cite{4415879} overlook design constraints in modern VLSI design and did not design the watermarking algorithms with the timing optimization objectives. It results in 1.51\% and 4.24\% WNSR degradation over the original designs, respectively. Furthermore, while Buffer Insertion~\cite{sun2006watermarking} encodes additional buffers on the non-timing critical path, the inserted buffers still result in 5.66\% TNSR and 10.63\% WNSR timing metrics degradation.

\subsubsection{Watermarking Capacity}
The watermarking capacity is measured by the maximum length of watermark bits that can be inserted into the layout without significantly degrading the layout quality. 
For Row Parity~\cite{kahng1998robust,kahng2001constraint}, Cell Scattering~\cite{4415879}, and \sys: DW, the signature length corresponds to the moved/inserted cell number. For Buffer Insertion~\cite{sun2006watermarking}, the length is the number of buffer-inserted nets. For \sys: GW, the signature length corresponds to the $N_w$ in Eqn.~\ref{eq:score}. For \sys, the signature length corresponds to the $N_w$ cells in global watermarking and moves $N_w$ cell over the watermarked region in the detailed watermarking. 

\begin{figure}[!ht]
    \centering
    \includegraphics[width= 0.95\columnwidth]{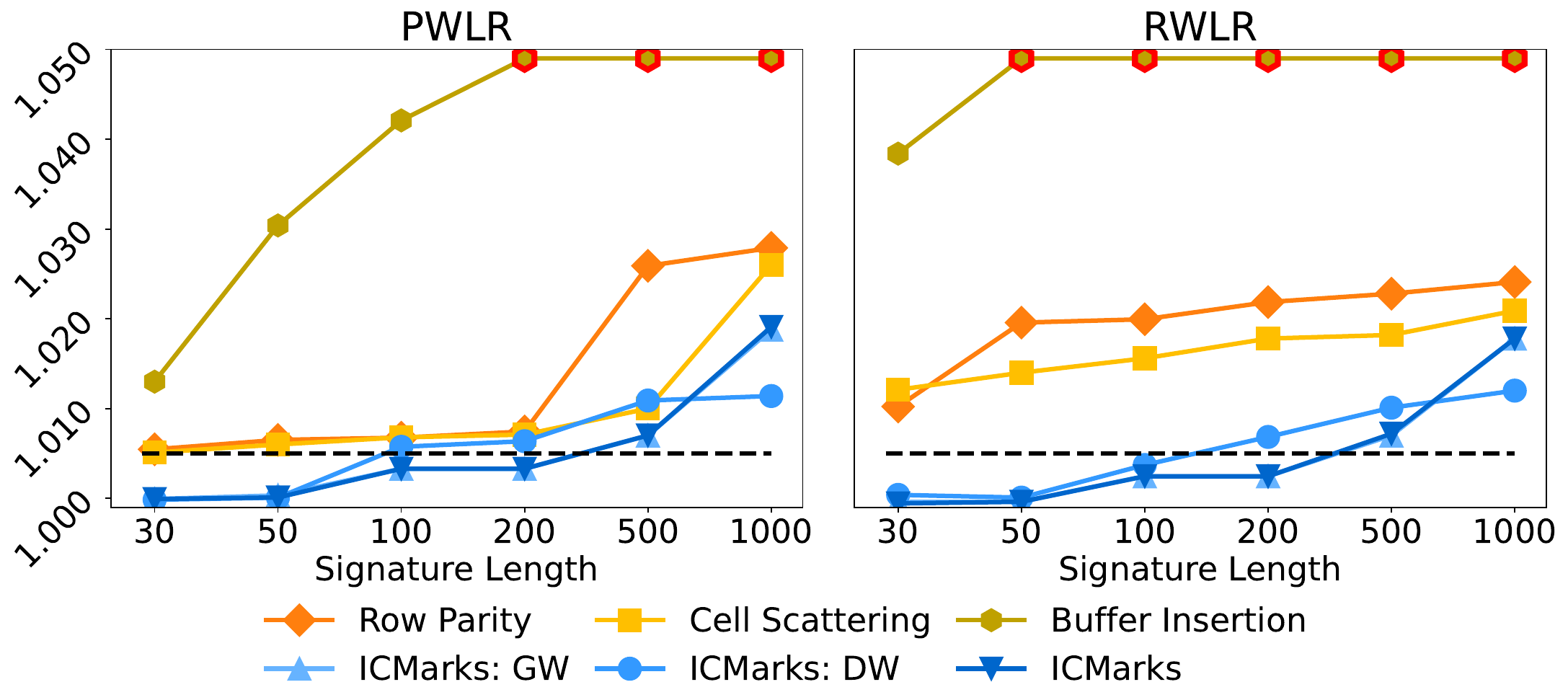}
     \caption{Watermarking capacity for different frameworks on ISPD'2019 benchmarks~\cite{liu2019ispd}. We consider the threshold for acceptable degradation layout quality as 0.5\%. The red $\redborderhexagon$ indicates PWLR and RWLR are higher than the 5\% limit. }
    \label{fig:capacity}
\end{figure}

We use ISPD'2019~\cite{liu2019ispd} as the benchmarking target and display the results in Fig.~\ref{fig:capacity}. As seen, the maximum bits that can be inserted by Row Parity and Cell Scattering are both less than 30 bits. Both methods move the cells across rows after the detailed placement as watermarks, which leads to worse RWLR than PWLR, as the router has to cross rows to connect the cells. 
Buffer Insertion~\cite{sun2006watermarking} inserts signatures without considering the design constraints, leading to significant cell displacement after signature insertion. Therefore, it exhibits low watermarking capacity on ISPD'2019~\cite{liu2019ispd}.
The \sys: DW's capacity is larger and reaches $\sim$ 100 bits. Because it moves the cells before detailed placement and the subsequent detailed placement compensates for the watermark insertion. \sys: GW and \sys{} further outperform the scheme. They both demonstrate a capacity of more than 200 bits by exploring the less compact region for watermark encoding.  Since \sys{} builds the watermarking scheme on top of \sys: GW, the two PWLR and RWLR lines are close in Fig.~\ref{fig:capacity}.

\begin{SCfigure}[50][t]
    \centering
    \includegraphics[width=0.5\linewidth]{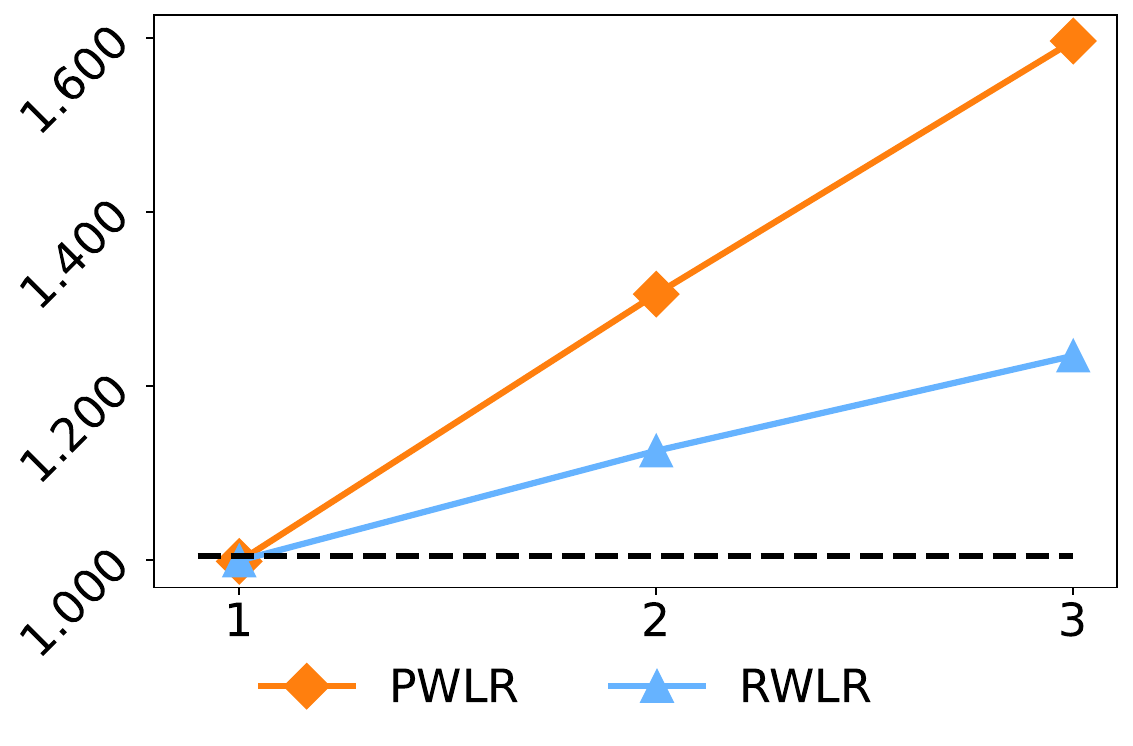}
    \vspace{-0.5cm}
    \caption{Watermarking capacity for watermark regions on ISPD'2019 benchmarks~\cite{liu2019ispd}.}
    \label{fig:capacity2}
\end{SCfigure}

\revision{Besides, we analyze the maximum number of watermark regions that can be encoded onto the layout without compromising the quality. We choose the best-scored $k$ regions in the GW stage and encode the $k$ regions during GW. As in Fig.~\ref{fig:capacity2}, encoding 3 watermark regions results in over 20\% wirelength degradation. When encoding two or more region constraints into global placement, more cells are on the selected watermark region boundary. Expelling the boundary cells to ensure the design adheres to watermarking constraints results in significant cell displacement and degrades the watermarked layout quality.} 

     
 
   

\subsubsection{Watermarking Efficiency}
The watermarking efficiency is benchmarked by the fraction of the additional time takes for \sys's watermark insertion compared to the overall physical design stage, primarily placement and routing.
We include the average time takes to encode 50-bit signatures onto the ISPD'2015 benchmark~\cite{bustany2015ispd} and ISPD'2019 benchmark~\cite{liu2019ispd} in Table~\ref{tab:watermark_overhead}. The non-WM time and WM time is the average (geometric-mean) time it takes for placement and routing on the non-watermarked and watermarked layout. The Slow Down is the percentage overhead \sys{} introduced to the overall placement and routing phase. As seen, the time taken for watermark insertion is  $\sim$10\% compared with the overall placement and routing stage, making \sys{} efficient for watermarking. Besides, no additional computation resources or external tools are required for the signature insertion.

%

\begin{table}[!h]
 \resizebox{0.95\columnwidth}{!}{%
\begin{tabular}{cccc}
\toprule
Design & non-WM Time (s)  & WM Time (s)  & Slow Down (\%) \\ \hline

ISPD'2015~\cite{bustany2015ispd} &  141.37 & 157.96  & 11.73\%\\
 ISPD'2019~\cite{liu2019ispd} & 946.23 & 996.18 & 9.21\%  \\
\bottomrule
\end{tabular}%
 }
\caption{\sys{}'s efficiency on different benchmarks. }
\label{tab:watermark_overhead}
\end{table}

\subsubsection{Watermarking Stealthiness}

\begin{figure}[!ht]
    \centering
    \vspace{-15pt}
    \subfloat[test5(WM)]{\label{f:design9} \includegraphics[width=0.24\columnwidth]{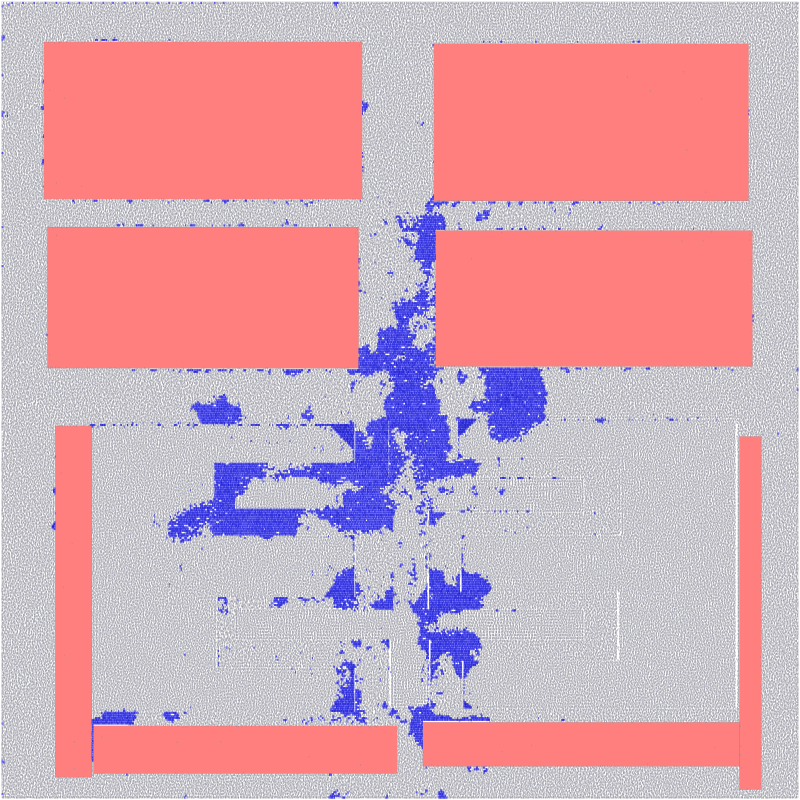}}
    \subfloat[test5(no-WM)]{\label{f:design9} \includegraphics[width=0.24\columnwidth]{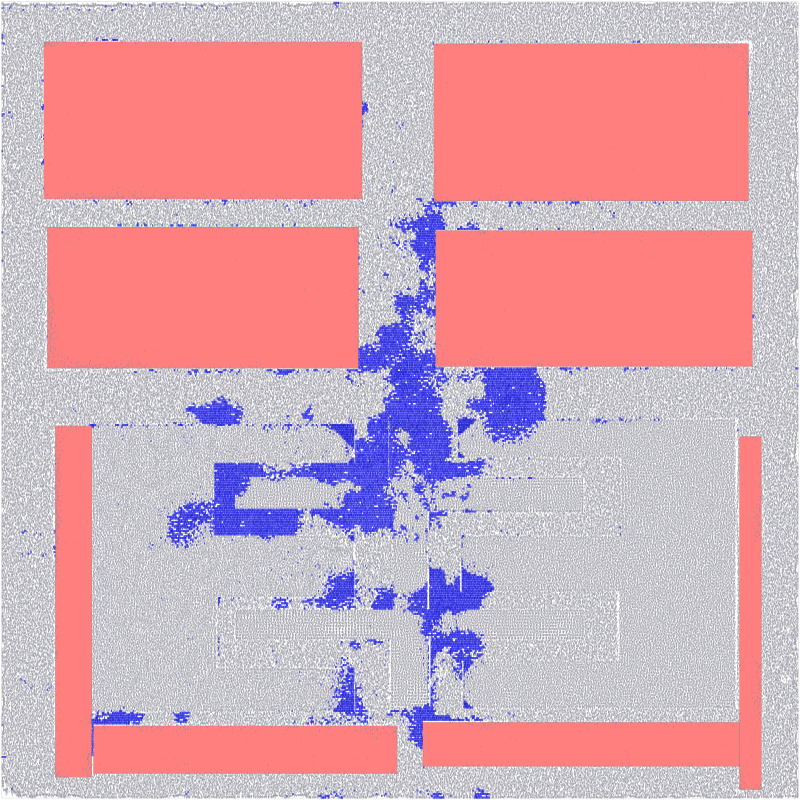}} 
    \subfloat[perf\_a(WM)]{\label{subf:design3} \includegraphics[width=0.24\columnwidth]{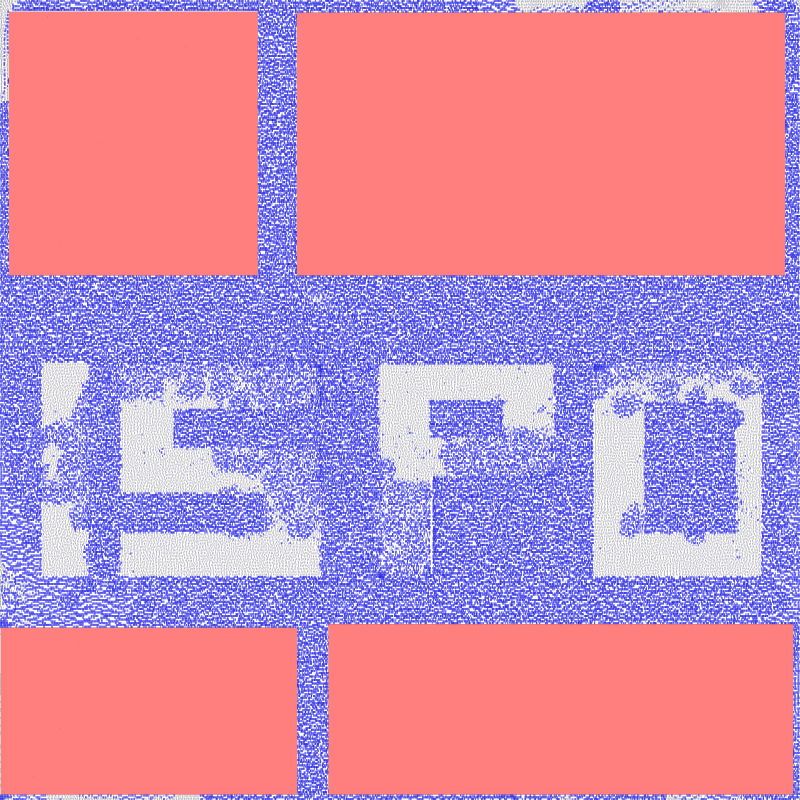}} 
   \subfloat[perf\_a(no-WM)]{\label{subf:design3} \includegraphics[width=0.24\columnwidth]{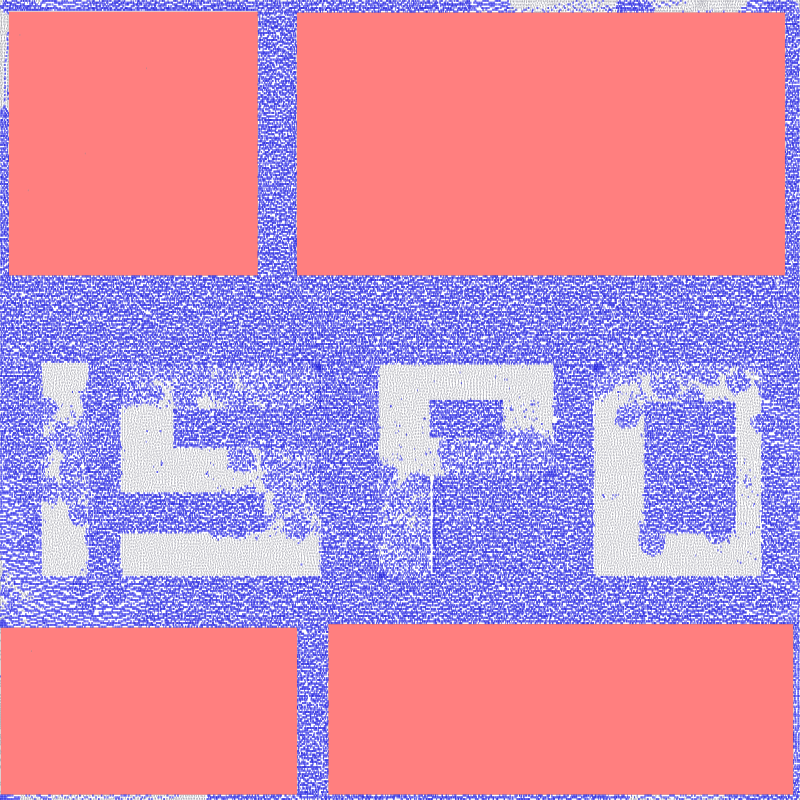}}
    \caption{Watermarked Design Examples. The \textcolor{blue}{blue} cells are the standard cells, and the \textcolor{red}{red} cells are the macros.}
    \vspace{-10pt}
    \label{fig:example}
\end{figure}

We display the layout watermarked by \sys{} with various sizes and different design constraints in Fig.~\ref{fig:example}.  As seen, the watermarks are embedded as part of the layout, and invisible upon inspection while maintaining 100\% WERs.  

\subsubsection{Watermark Strength} \revision{Watermark strength measures the probability a non-watermarked layout carries the signature by coincidence. For each design with a total of $|C|$ cells, we watermark $|C_w|$ cells and produce the watermarked layout. In the $|C_w|$ cells, $x$ of the cells do not meet the watermarking constraints.  
$|C_n|$ from the non-watermarked layout matches the watermark requirement by co-incidence. Following Row Parity~\cite{kahng2001constraint}, the watermark strength is calculated by Eqn.~\ref{eq:strength} as follows, where $p=\frac{|C_n|}{|C_w|}$ is the constraints met by coincidence.} 

\begin{equation}
    \label{eq:strength}
    P_c=\sum_{i=0}^{x}\cdot \mathbf{C}(|C_w|, i)  \cdot(p)^i \cdot (1-p)^{x-i}
\end{equation}

\revision{For each design, we calculate the $P_c$ by running the placement with and without \sys{} to obtain $|C_w|$ and $|C_n|$. In most of the designs, $|C_n|=0$, meaning the non-watermarked layouts do not match the watermarking constraints. Therefore, we report the maximum $P_c$ on ISPD'2019 benchmark~\cite{liu2019ispd} in Table~\ref{tab:bits} for \sys{} and its submodules.} \revisionsecond{The watermarks are successfully encoded during the global/detailed placement stage and have different positions from their non-watermarked ones. As such, the signatures' randomness will not be undermined by the global/detailed placement optimizations. }

\begin{table}[ht]
  \resizebox{0.95\columnwidth}{!}{
     \centering
    \small
    \begin{tabular}{ccccc}
    \toprule
        & Row Parity & \sys: GW & \sys: DW & \sys: GW + DW \\
        \hline
     $P_c$ & 8.8$\times10^{-16}$ & 9.09 $\times10^{-53}$ &  8.08$\times10^{-62}$ & 7.35$\times10^{-114}$\\
    \bottomrule
    \end{tabular}}
    \caption{Watermarking strength for different signature lengths by the proposed \sys{} watermarking framework.}
     \label{tab:bits}
\end{table}

\subsubsection{Ownership Proof in Real-World Settings} To prove ownership, the design company obtains the suspicious layout and employs reverse-engineering approaches~\cite{9300272,alrahis2021gnn} to acquire the logic netlist and all standard cell locations. Such methodology~\cite{9300272,alrahis2021gnn} recovers large layouts netlists (over 7000k cells) from the GDSII layout with over 98\% accuracy and efficiency. Then, the design company uses the netlists to recover the standard cells' and macros' locations.  The watermark extraction algorithms in Section~\ref{sec:method} are subsequently used for ownership proof. 
While the reverse-engineering misalignment might degrade the watermark extraction rates slightly, \sys{} still provides sufficient ownership proof for the design companies benefiting from two aspects: (i) high watermarking strength of 7.35$\times10^{-114}$ for 50-bit signature; (ii) high watermarking capacity that can accommodate more than 200-bit signatures without significant quality degradation, as in Fig.~\ref{fig:capacity}. 
As a result, even if only 88\% signatures are extracted (10\% lost from attacks in Section~\ref{sec:security_ana} and 2\% lost from reverse engineering~\cite{9300272,alrahis2021gnn}), \sys{} still provides strong watermarks for 50-bit signatures. 
In Null Hypothesis~\cite{anderson2000null}, a watermarking strength (p-value) of smaller than 0.05 indicates the statistically significant presence of a watermark. Given the much lower watermarking strength than 0.05 after reverse engineering, the design company can thereby confidently claim ownership of the design layout. 

\subsubsection{Impact on Routability} 
\begin{figure}[!ht]
    \centering
    \vspace{-8pt}
    \includegraphics[width=0.85\columnwidth]{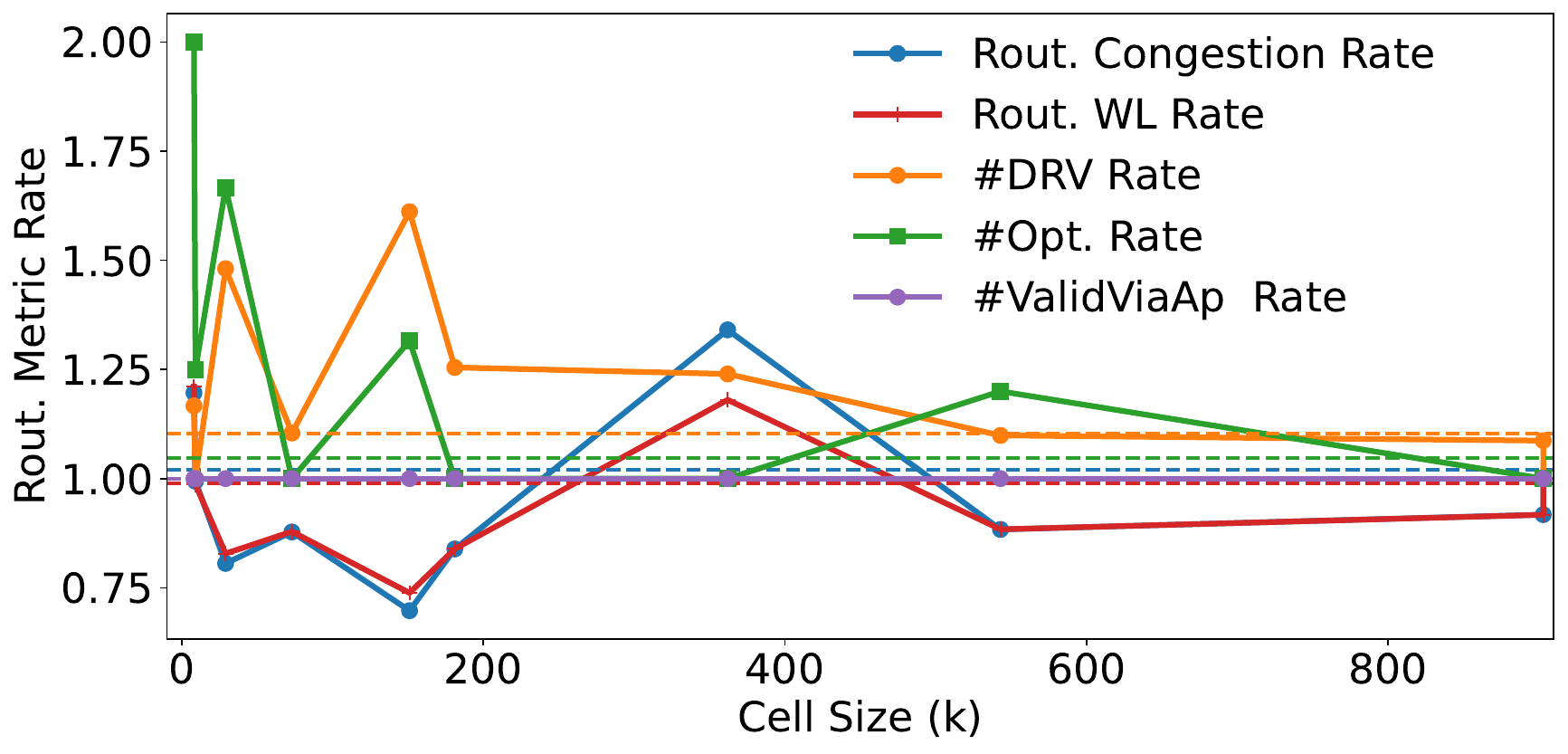}
    \caption{Impact of watermarking on routability. The dotted lines represent the geo-mean of large designs ispd19test7 - ispd19test10 with $> 300k$ cells.
 }
    \vspace{-20pt}
    \label{fig:route_metrics}
\end{figure}

\revisionsecond{We analyze the impact of \sys{} on the design routability using the ISPD'2019~\cite{liu2019ispd} benchmarks employing OpenROAD~\cite{kahng2021openroad}'s global and detailed router.
The final wirelength after detailed routing (Rout. WL), average routing congestion (Rout. Congestion), an initial count of DRC violations (\#DRV), the number of detailed routing optimization iterations to fix DRV (\#Opt.), and the total valid via access points required for all standard cells and macros in the design (\#ValidViaAp) are plotted in Fig.~\ref{fig:route_metrics} for the watermarked layouts over the non-watermarked ones.
}

\revisionsecond{
Across all the considered benchmarks, the watermarked ones have $6\%$ lower Rout. Congestion and Rout. WL compared to the non-watermarked designs.
For large layouts with $ > 300k$ cells, 
\sys{} introduces $2\%$ more routing congestion than non-watermarked ones, showing it has minimal impact on layout routability.
Nevertheless, \sys{} does incur an increase in \#DRV and \#Opt., by $19\%$ and $20\%$, respectively, across all the designs for watermarked ones, and $10\%$ more \#DRV requiring $\sim 5\%$ more \#Opt. for the large designs ispdtest7 - ispdtest10 consisting of $> 300k$ cells.
\#ValidViaAp, representing the layout's pin accessibility, remains similar for watermarked and non-watermarked designs.
Thus, \sys{} does not degrade the pin access of the design. 
}

\section{Attack Evaluation}\label{sec:security_ana} 

In this section, we evaluate \sys's robustness under watermark removal and forging attacks. 

\subsubsection{Watermark Removal Attack}\label{subsec:wm_removal}
\begin{figure*}[!ht]
    \centering
    \subfloat[Swap location attack (0.1\%) ]{\label{f:sla_10_attack} \includegraphics[width=0.5\linewidth]{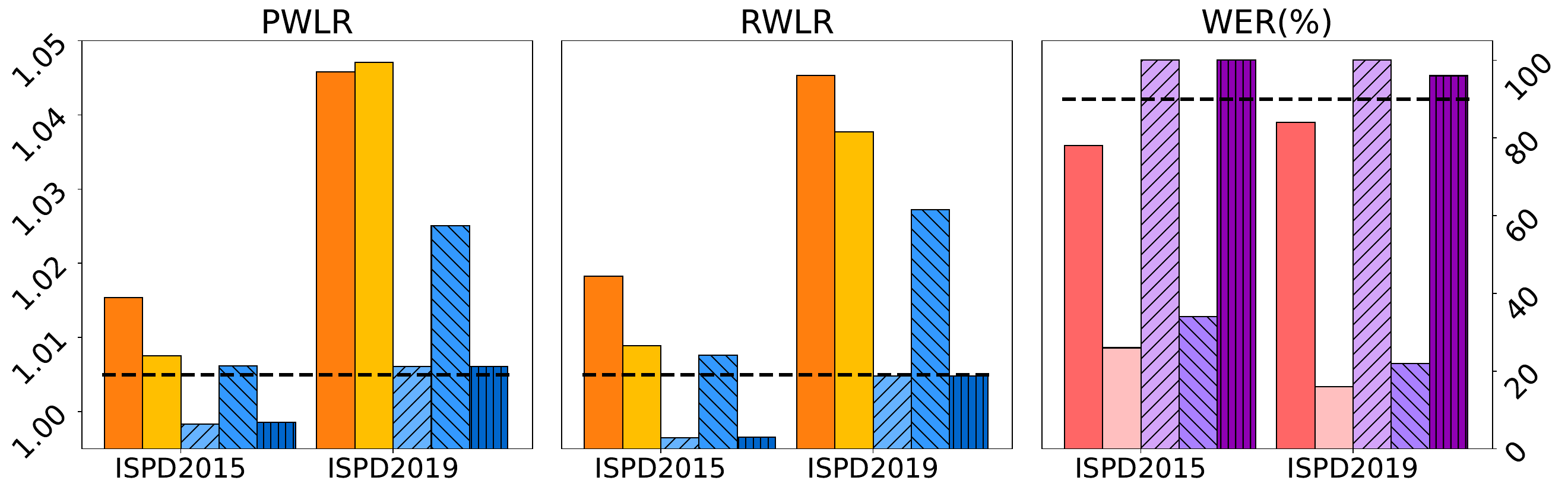}}
    \subfloat[Swap location attack (0.5\%) ]{\label{f:sla_50_attack} \includegraphics[width=0.5\linewidth]{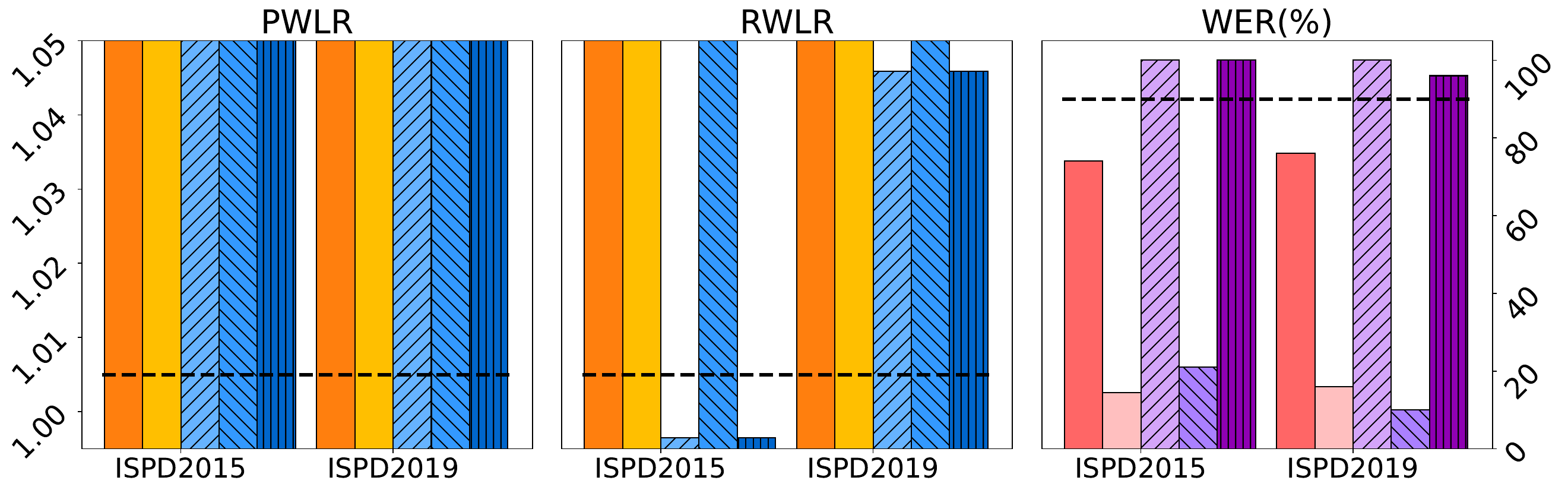}} \\
    \subfloat[Constraint perturbation attack (0.1\%) ]{\label{f:cpa_0.001} \includegraphics[width=0.5\linewidth]{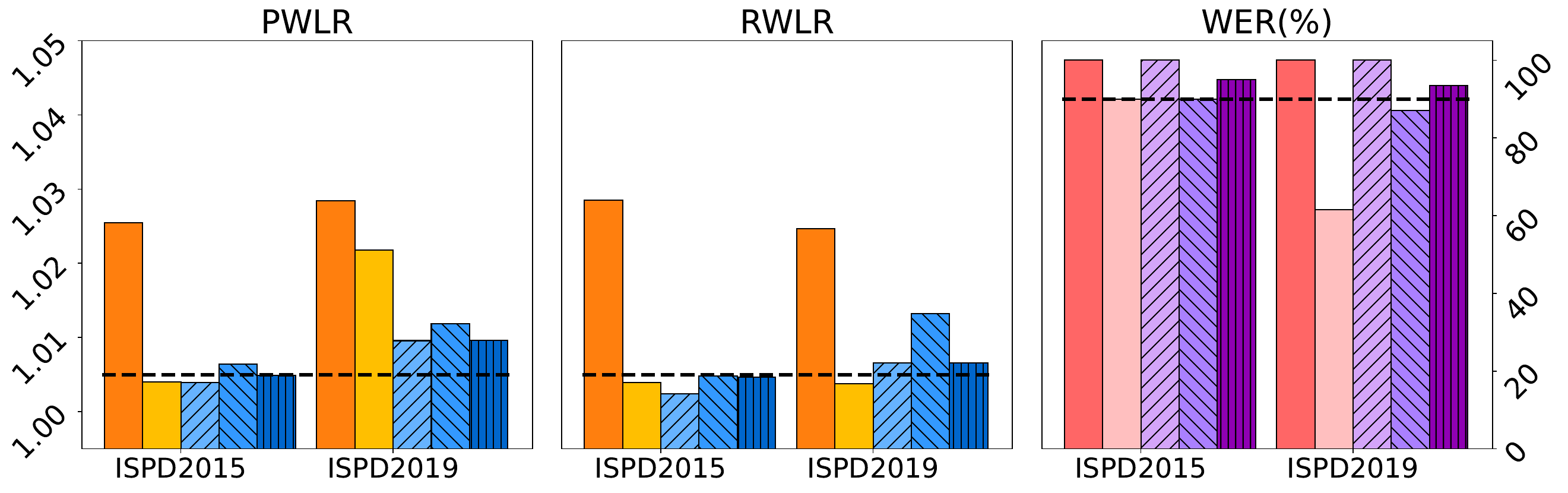}} 
    \subfloat[Constraint perturbation attack (1\%) ]{\label{f:cpa_0.01} \includegraphics[width=0.5\linewidth]{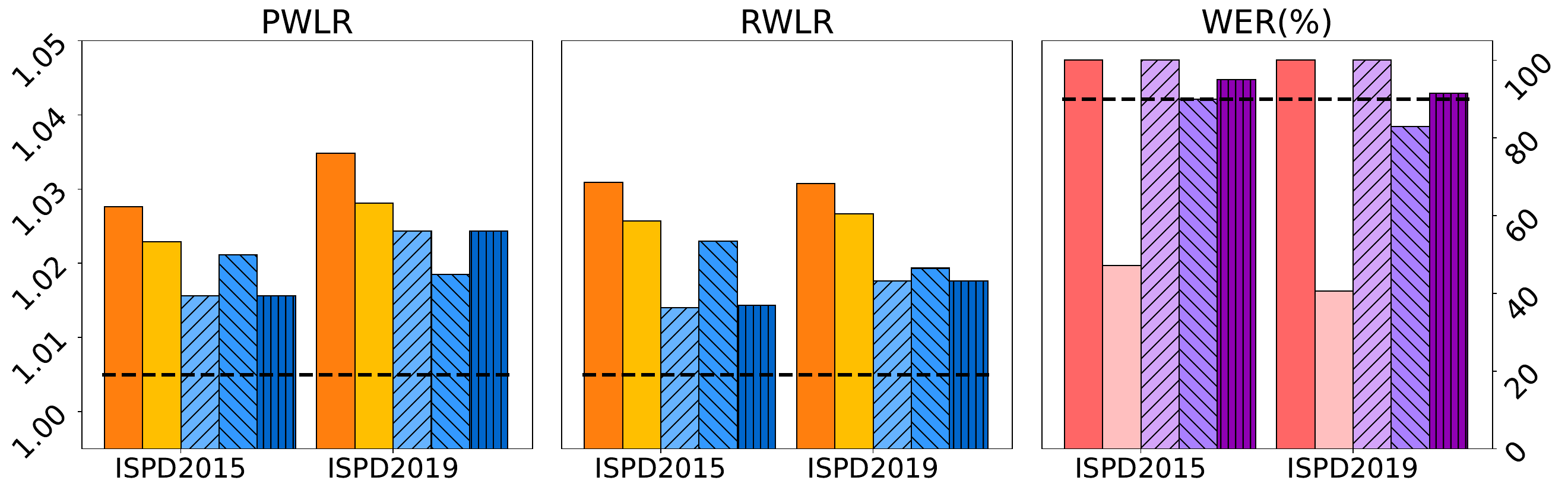}} \\ 
    \subfloat[Constraint perturbation attack (10\%)]{\label{f:cpa_0.1} \includegraphics[width=0.5\linewidth]{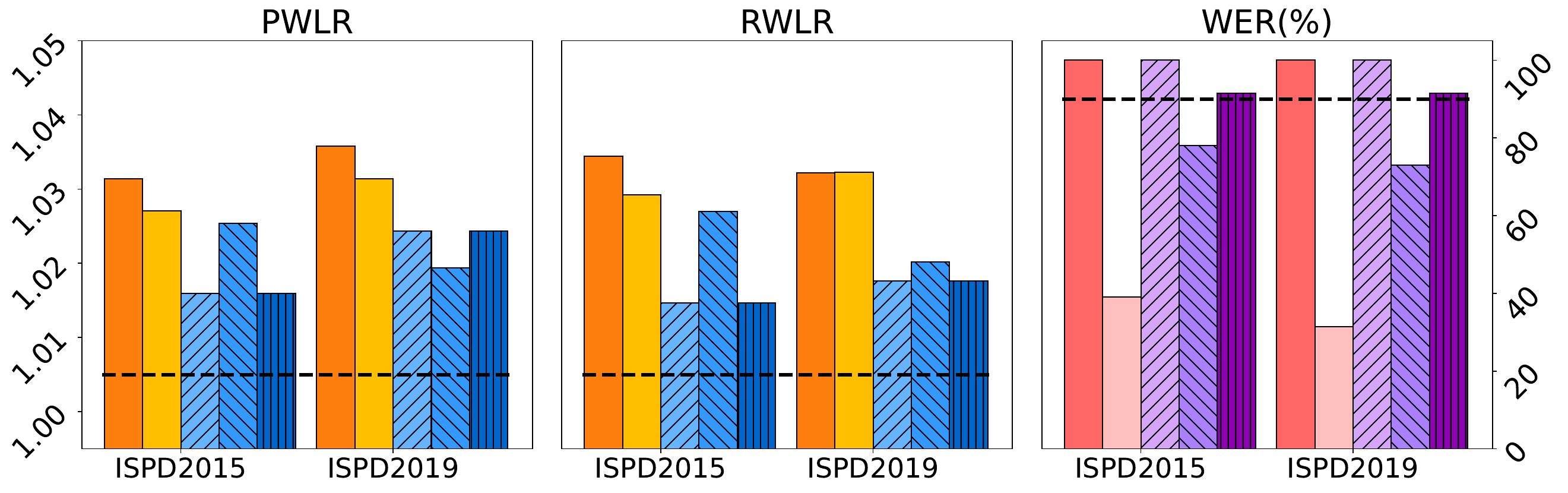}} 
    \subfloat[Optimization attack ]{\label{f:eco_attack} 
    \includegraphics[width=0.5\linewidth]{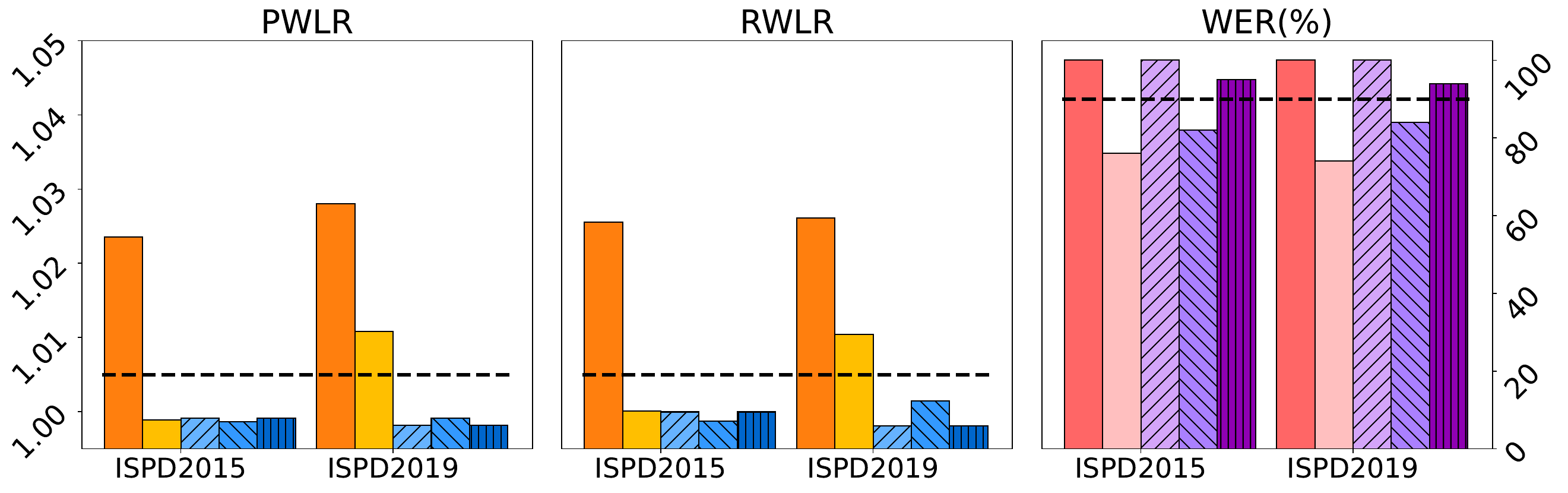}} \\ 
    \subfloat[Adaptive region attack (top-1) ]{\label{f:ada_attack1} \includegraphics[width=0.5\linewidth]{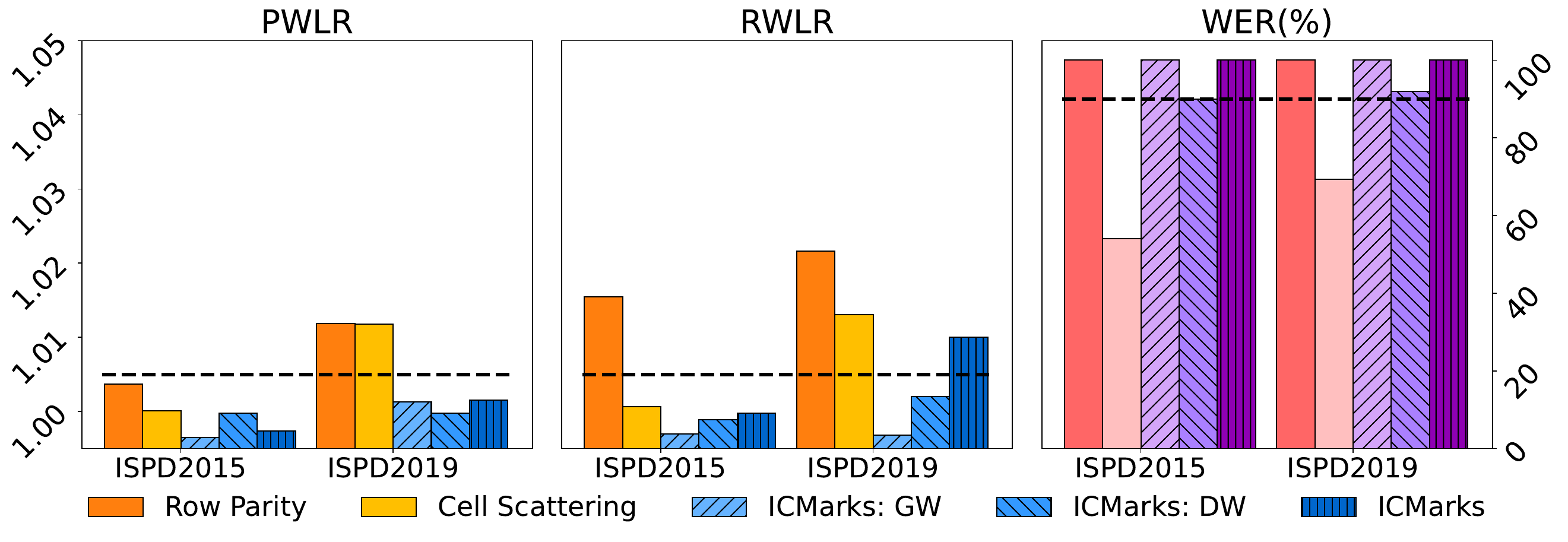}}
   \subfloat[Adaptive region attack (top-5) ]{\label{f:ada_attack5} \includegraphics[width=0.5\linewidth]{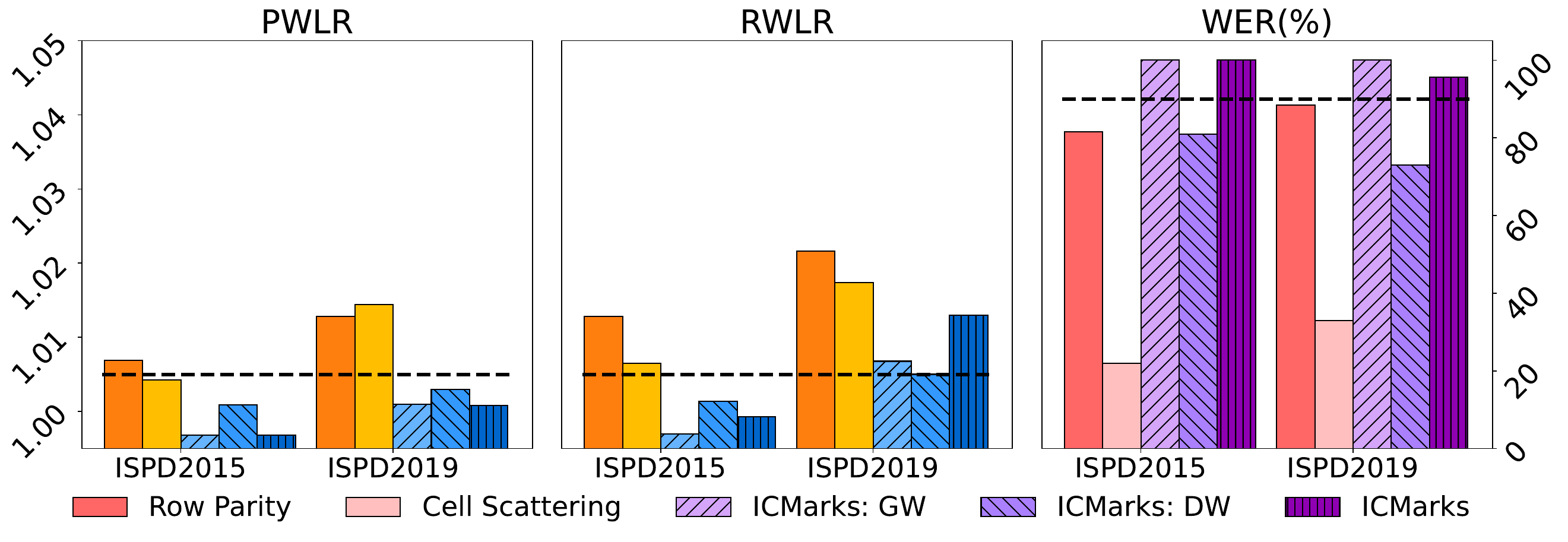}}
    \caption{Watermarking performance under different attacks for wirelength-driven placement on the ISPD'2015~\cite{bustany2015ispd} and ISPD'2019~\cite{liu2019ispd} benchmarks. The black dotted line in the two left subfigures denotes the quality degradation threshold of 1.005, and the black dotted line in the rightmost subfigure denotes the watermark extraction threshold of 90\%.}
    \vspace{-10pt}
    \label{fig:attack_results}
\end{figure*}

\begin{figure*}[!ht]
    \centering
    \subfloat[Swap location attack (0.1\%)]{\label{f:append_sla_10_attack} \includegraphics[width=0.25\linewidth]{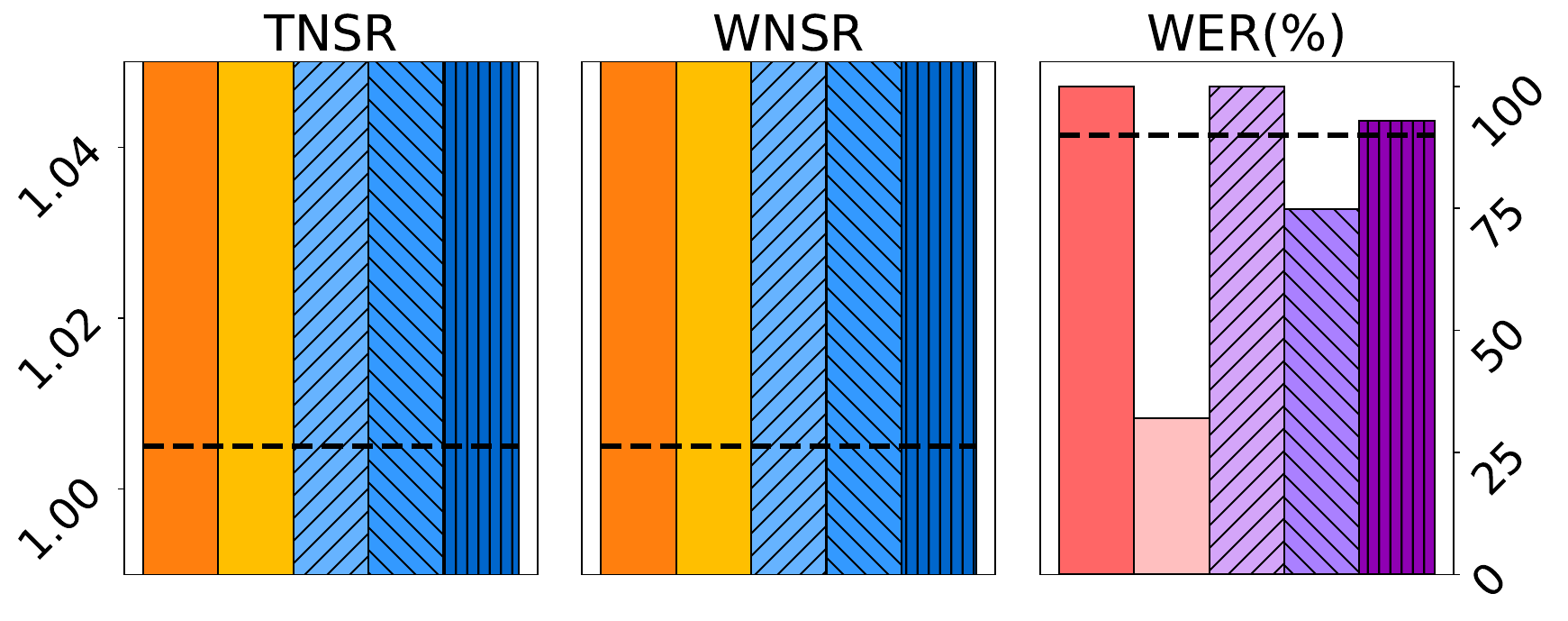}}
    \subfloat[Swap location attack (0.5\%)]{\label{f:append_sla_50_attack} \includegraphics[width=0.25\linewidth]{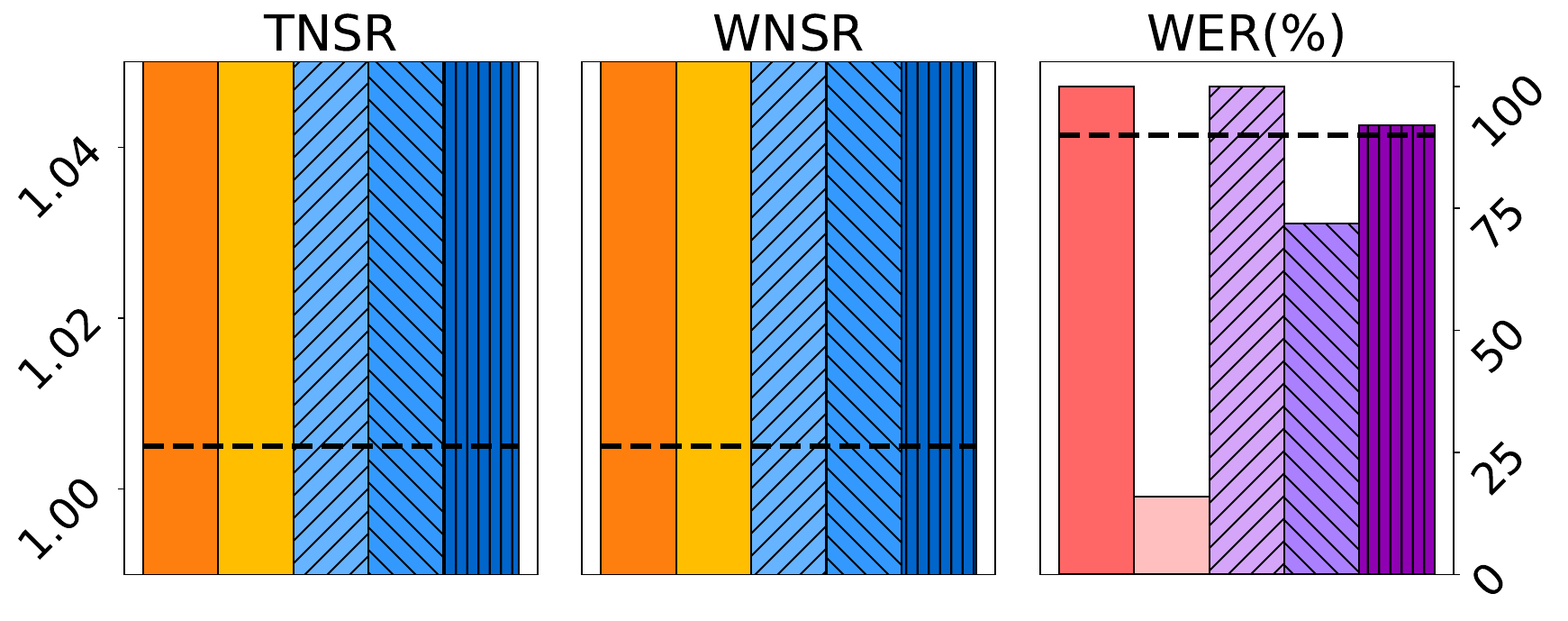}} 
    \subfloat[Constraint perturbation attack (0.1\%)  ]{\label{f:append_cpa_0.001} \includegraphics[width=0.25\linewidth]{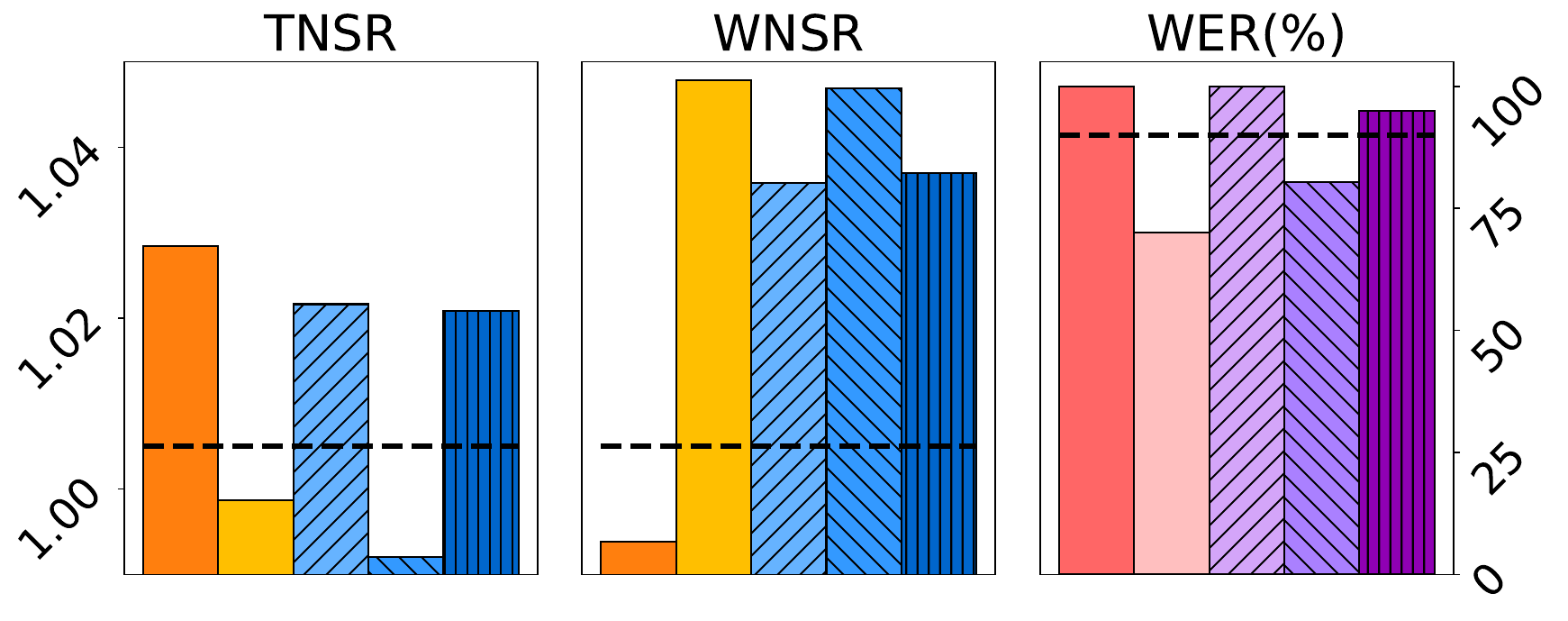}} 
    \subfloat[Constraint perturbation attack (1\%) ]{\label{f:append_cpa_0.01} \includegraphics[width=0.25\linewidth]{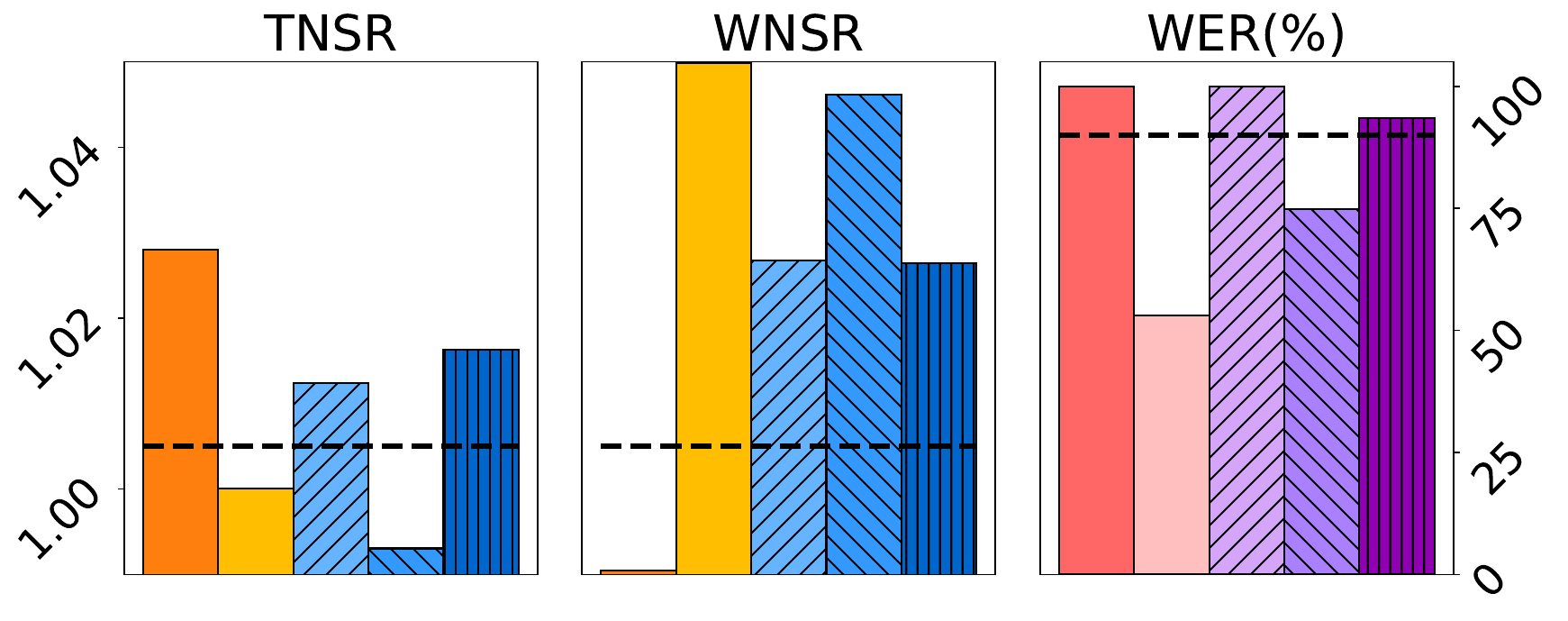}} \\
    \subfloat[Constraint perturbation attack (10\%)]{\label{f:append_cpa_0.1} \includegraphics[width=0.25\linewidth]{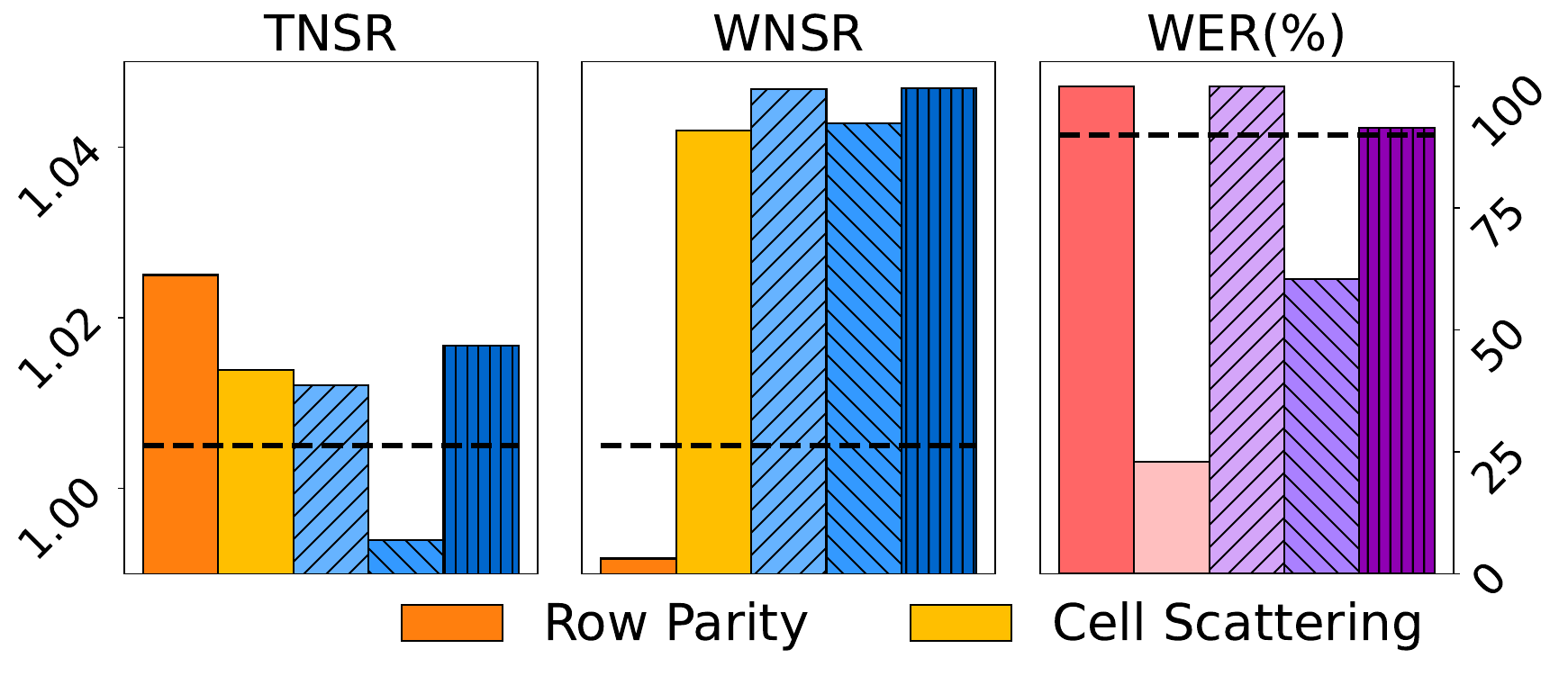}} 
    \subfloat[Optimization attack ]{\label{f:append_eco_attack} 
    \includegraphics[width=0.25\linewidth]{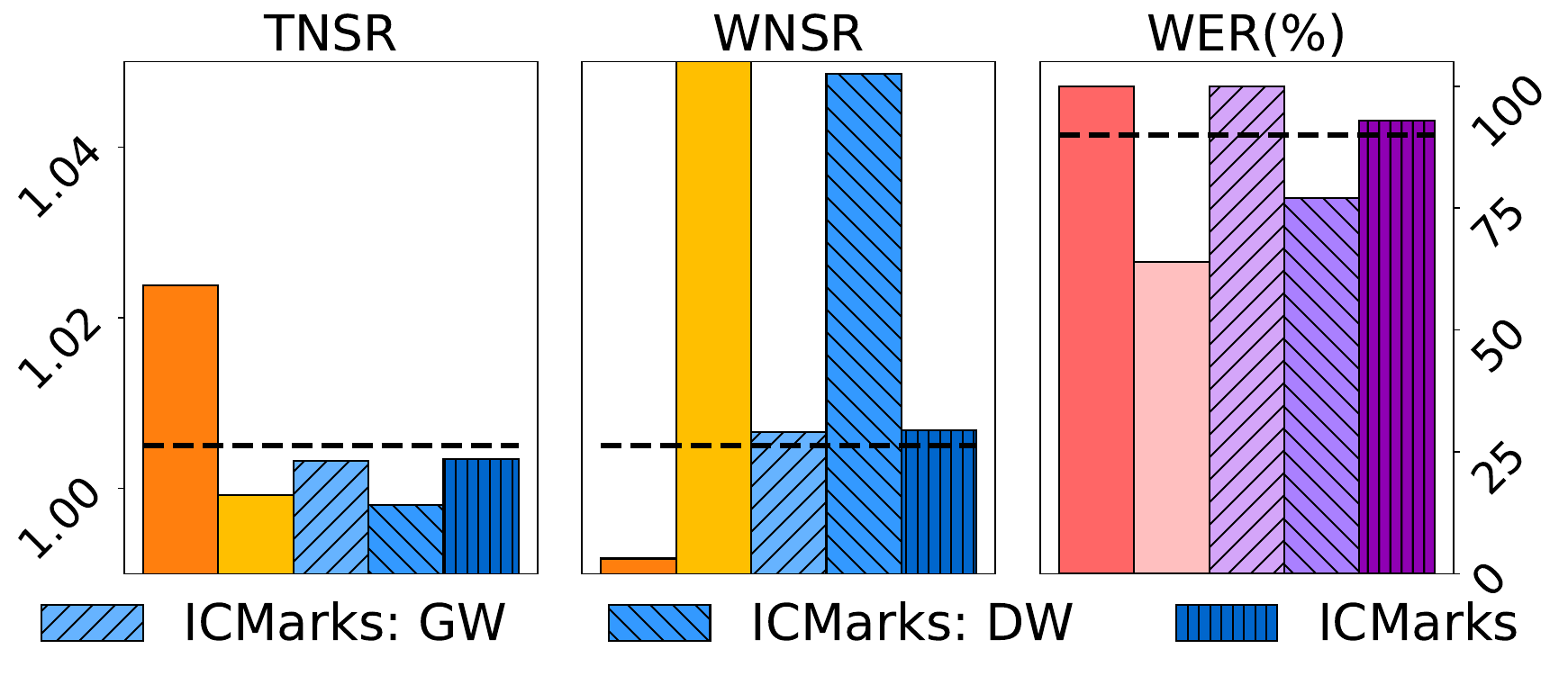}}   
    \subfloat[Adaptive region attack (top-1) ]{\label{f:append_ada_attack} \includegraphics[width=0.25\linewidth]{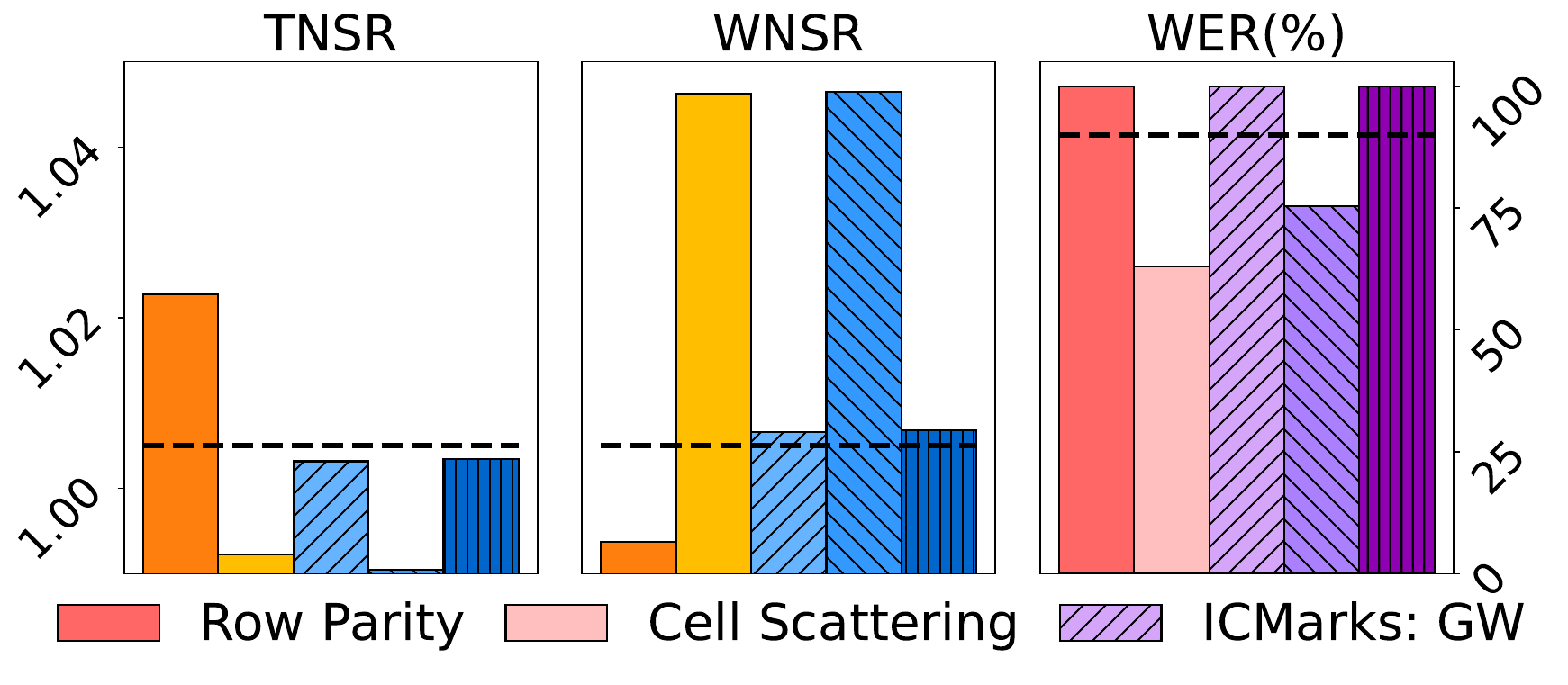}}
   \subfloat[Adaptive region attack (top-5)]{\label{f:ada_attack} \includegraphics[width=0.25\linewidth]{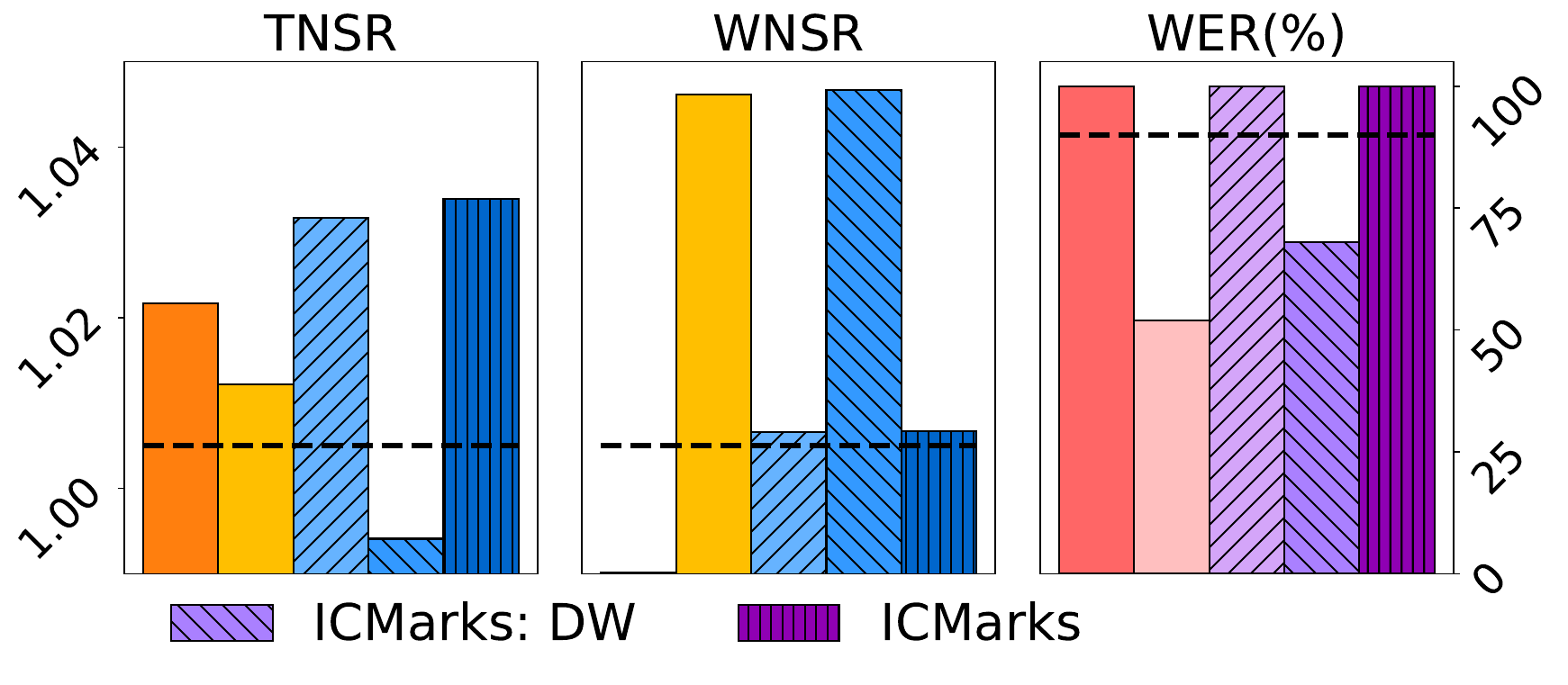}}
    \caption{Watermarking performance under different attacks for timing-driven placement on the ICCAD'2015 benchmarks~\cite{7372671}. The black dotted line in the two left subfigures denotes the quality degradation threshold of 1.005, and the black dotted line in the rightmost subfigure denotes the watermark extraction threshold of 90\%.}
\vspace{-10pt}
    \label{fig:timing_attack_results}
\end{figure*}

A successful watermark removal attack shall meet two criteria: (i) the encoded signatures are removed as reflected by the watermark extraction rate (WER) below 90\%~\cite{kahng1998robust,kahng2001constraint}; (ii) the layout quality is not compromised, where the Placement Half-Perimeter WireLength Rate (PWLR), Routed Wirelength Rate (RWLR), Total Negative Slack Rate (TNSR), and Worst Negative Slack Rate (WNSR) do not exceed 1.005~\cite{qiu2023progress}. The thresholds are reflected by the black dotted lines in Fig.~\ref{fig:attack_results} and Fig.~\ref{fig:timing_attack_results}.




For an IC layout watermarked by \sys{}, the watermarks are inserted by constraining the cell positions and cells' region. To remove the signature, the adversary perturbs the cell positions/regions for watermark removal attacks. 
We compare four types of attacks aiming to erase the signatures at different levels of watermarking and prevent the owner from claiming ownership: (i) swap location attacks (SLA) ~\cite{kahng2001constraint}, which target to attack Row Parity framework by random swapping cell locations, (ii) constraint perturbation attacks (CPA), which target to attack Cell Scattering and \sys: DW by moving cells along x/y axis if there is space; (iii) Optimization attacks (OA), which target to attack all watermarking frameworks by running another round of detailed placement; and (iv) adaptive region attacks (ARA), which target to attack \sys: GW and \sys{} framework by searching for less compacted regions and perturb cells around the region. We show \sys{} is robust against all removal attacks, and the results are summarized in Fig.~\ref{fig:attack_results} and Fig.~\ref{fig:timing_attack_results} for wirelength-driven and timing-driven placements respectively.


We skip Buffer Insertion~\cite{sun2006watermarking} and the corresponding watermark removal attacks because (i) Buffer Insertion's~\cite{sun2006watermarking} watermarked layout quality degradation is significantly higher than the considered threshold of 0.5\% as in Table~\ref{tab:ispd2015}-\ref{tab:iccad2015}; (ii) Other watermarking frameworks insert signatures without modifying the netlist, whereas Buffer Insertion~\cite{sun2006watermarking} encodes adding buffers into the netlist as watermarks. While the buffer removal attacks, which remove multiple cascaded buffers, can potentially erase Buffer Insertion's signature, our primary goal is evaluating the robustness of \sys{} under attacks. Therefore, we did not include the attacks at the netlist level in this section. 





\textbf{Swap Location Attacks (SLA)}:
In SLA~\cite{kahng2001constraint}, cells randomly exchange their locations with another set of cells. Then, a follow-up legalization and detailed placement compensate for the performance degradation and ensure the cells follow the design rules. 
In Fig.~\ref{f:sla_10_attack} and Fig.~\ref{f:sla_50_attack}, we randomly choose 0.1\% and 0.5\% cells from the layout and pair them to exchange their locations. As seen, even small location swaps result in huge performance degradation on modern IC layouts. In terms of the watermark extraction, Row Parity~\cite{kahng1998robust,kahng2001constraint}, Cell Scattering~\cite{4415879}, and \sys: DW get WERs below 90\% and failed to verify their ownership. Those watermarking frameworks spread the watermarks across the layout, where minor changes will subsequently modify most cell locations within the compact region. 
By embedding cells in the less compact region, \sys: GW and \sys{} are less sensitive to such changes.  
As a result, \sys: GW and \sys{} still maintain high WERs.

\textbf{Constraint Perturbation Attacks (CPA)}:
In CPA, cells shift their location along the x-axis for $\delta_x = 1$ or y-axis for $\delta_y$ set to one adjacent row-height if such movements do not result in overlapping with their neighbors. In Fig.~\ref{f:cpa_0.001}, Fig.~\ref{f:cpa_0.01}, and Fig.~\ref{f:cpa_0.1}, we move the positions of 0.1\%, 1\%, 10\% of the cells have space to move in the layout. 
From here, we find that under CPA 0.1\% attack, layout performance hit the boundary of the wirelength degradation (PWLR and RWLR) threshold. For the watermark extraction, \sys{}, \sys: GW, and Row Parity~\cite{kahng1998robust,kahng2001constraint} achieve over 90\% WER. However, the WER of Cell Scattering~\cite{4415879} and \sys: DW are subsequently lower than 90\% because the constraint perturbation targets to remove the potential watermarked cell positions. 

For the CPA 1\% and 10\% cell performance, however, the PWLR and RWLR degradations are greater than 1.005, meaning the attack failed to maintain the layout quality improvement from the design company's physical design optimizations. But \sys: GW and \sys{} still maintain higher than 90\% WER and successfully help the design company to claim ownership of the layout.

\textbf{Optimization Attacks (OA)}:
This attack employs an additional optimization stage to remove the watermarks.
The optimization is implemented through another round of detailed placement. It aims to change cell locations slightly for signature removal while maintaining the layout quality. As shown in Fig.~\ref{f:eco_attack}, the layout wirelength degradation in Cell Scattering~\cite{4415879}, \sys: GW, \sys: DW, and \sys{} are all below 1.005, indicating the attack preserves the layout quality. The WER of Cell Scattering~\cite{4415879} and \sys: DW are below 90\%, meaning OA successfully removed their signatures. In contrast, \sys: GW and \sys{} have over 90\% WER, demonstrating their resiliency.



\textbf{Adaptive Region Attacks (ARA)}:
This attack targets to remove watermarks in \sys:GW. The adversary has prior knowledge of how \sys{}:GW performs the watermarking and has access to the hyperparameters used to search the watermarked region. The adversary operates on top of the watermarked layout.
He tries to remove the inserted watermarks by moving cells within the searched top-1 or top-5 regions if there is room.
In Fig.~\ref{f:ada_attack1} and Fig.~\ref{f:ada_attack5}, the PWLR and RWLR degradation for \sys: GW, \sys: DW, and \sys{} are around the threshold of 0.5\% degradation, and the attacks do not significantly degrade the layout quality. The WER of \sys: GW and \sys{} remain over 90\%. As the watermark insertion is performed on a non-watermarked layout, and the attack regions searched by ARA are on a watermarked layout, the watermark signatures are not the same. Therefore, ARA fails to remove \sys: GW and \sys's signature, even if the same region watermarking algorithm is employed. In contrast, the watermarks of Row Parity~\cite{kahng1998robust,kahng2001constraint}, Cell Scattering~\cite{4415879}, and \sys:DW are spread across the layout, where minor changes in the compact area will subsequently modify the inserted watermarks. Therefore, these frameworks have compromised WER.

\textbf{Attacks on Timing-driven Placement}: In Fig.~\ref{fig:timing_attack_results}, the attack performance on timing-driven placement follows a similar trend as the wirelength-driven placement results in Fig.~\ref{fig:attack_results}. \sys{} and \sys: GW remain resilient under all the removal attacks with a WER over 90\%. Row Parity~\cite{kahng1998robust,kahng2001constraint} is also resilient to the attacks and has WER of over 90\%. However, the watermarked layouts degrade the timing metrics (TNSR and WNSR) by $\ge$ 1\% in Table~\ref{tab:iccad2015} for timing-driven placement. The signatures in the baseline Cell Scattering~\cite{4415879} and \sys:DW are removed with WER $<$ 90\%. 
%
%

\subsubsection{Watermark Forging Attack}
Instead of removing the watermarked signature, the adversary in a watermark forging attack counterfeits another set of watermarks on the watermarked layout and falsely claims his ownership. Row Parity~\cite{kahng1998robust,kahng2001constraint} and Buffer Insertion~\cite{sun2006watermarking} techniques are not resilient to forging attacks. If the adversary has prior knowledge of the watermarking algorithm, they can easily counterfeit a different set of forged signatures from the row index ID of cells or add additional buffers into the layout for false ownership proof. For Cell Scattering~\cite{4415879}, signatures are harder to counterfeit because forging the signature requires both random seeds and a non-watermarked layout. However, the inserted signatures can be easily erased by different watermark removal techniques in Section~\ref{subsec:wm_removal}.

\sys: GW is resilient to forging attacks because the region with the minimal evaluation score is unique to the original non-watermarked placement. The non-watermarked layout and the scoring parameters are kept confidential. The \sys: GW signature verification requires the owner to provide that information to reproduce the watermark region $R_w$. Therefore,
the adversary cannot counterfeit the watermarks without access to the non-watermark layout. \sys: DW also exhibits similar properties, where the signatures are encoded before detailed placement on the intermediate placement $P_{itr}$. As such, the adversary with only access to the watermarked layout cannot reproduce the watermarked cells or distance to forge the signature.  \sys{}, as a combination of \sys: GW and \sys: DW is also resilient to watermark forging attacks. 

\revision{We evaluate the probability that an adversary forges the \sys: GW signature, and we assume the adversary has prior knowledge of the scoring mechanism and watermarked region size. However, he does not have access to the scoring hyperparameters. Different from the Adaptive Region Attacks (ARA) in Watermark Removal Attacks, the adversary forges the signature by moving the cells in the searched region. We re-score the region using hyperparameters $\alpha = 25$, $\beta = 15$, and $\gamma = 10$.}

\begin{SCfigure}[50][ht]
    \centering
    \includegraphics[width=0.5\linewidth]{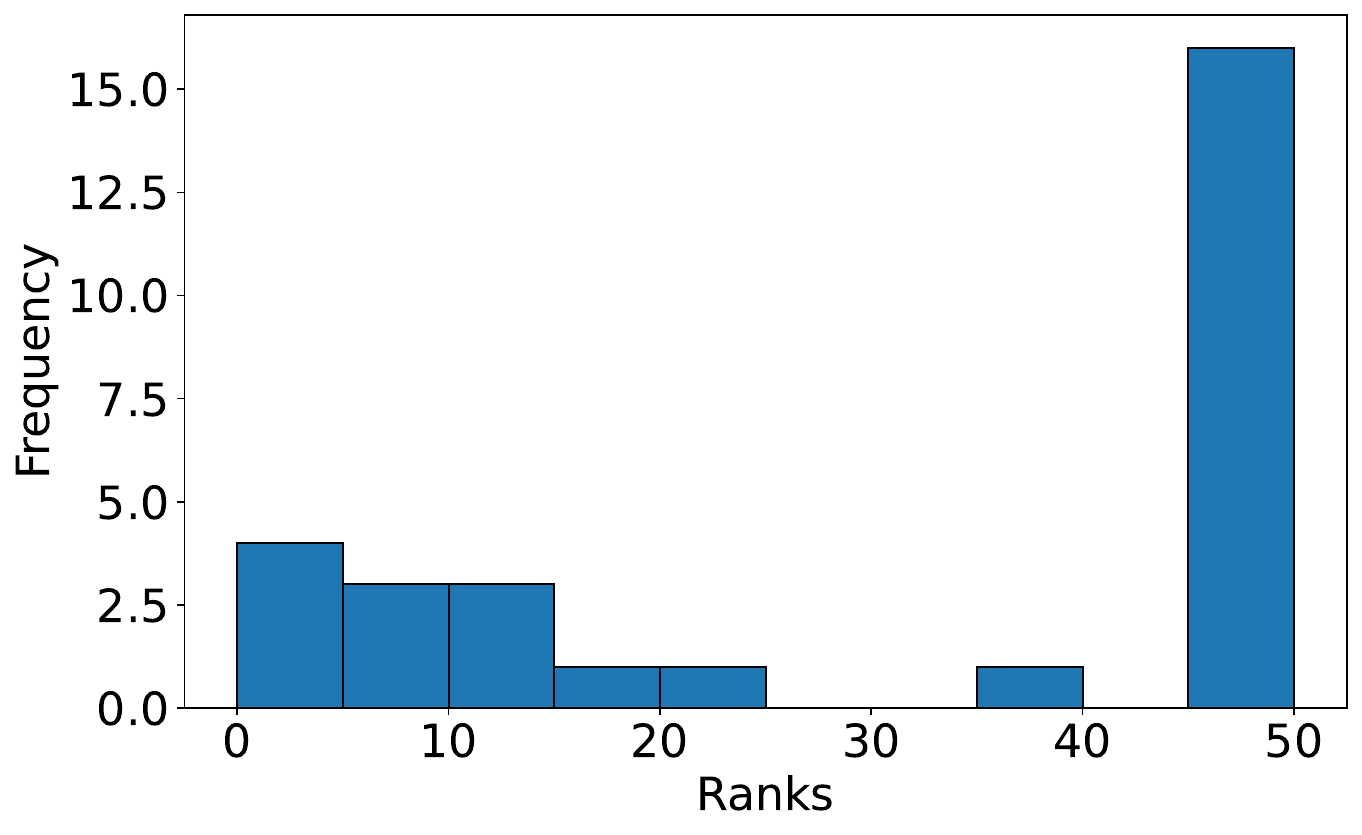}
    \vspace{-0.5cm}
    \caption{Rank distribution of the watermarked region appears in the re-scored watermarked region.}
    \label{fig:rank}
\end{SCfigure}

\revision{We rank the scores the adversary gets from low to high, where lower scores indicate the region is more ideal for him/her to forge the signature. 
We show the rank of the watermarked region appears in the re-scored watermarked layout in Fig.~\ref{fig:rank} on both ISPD'2019~\cite{liu2019ispd} and ISPD'2015~\cite{bustany2015ispd} benchmarks. As seen, most of the layouts are ranked away from the top-5 regions, meaning the adversary cannot easily forge the watermarked region of cells.}

\section{Conclusion}
\label{sec:conclusion}
We present \sys{}, a robust watermarking framework for integrated circuits physical design IP protection in the supply chain. We first introduce a \textit{Global Watermarking} method that identifies the watermarked region with insignificant performance degradation; then, we propose an independent \textit{Detailed Watermarking} technique to select cells that do not overlap with neighbors after perturbation to encode watermarks. Based on these methods, we develop \sys{}, which combines the best attributes of both \textit{Global} and \textit{Detailed Watermarking}, thereby achieving minimal quality degradation with augmented robustness. Extensive experiments on ISPD'2015~\cite{bustany2015ispd}, ISPD'2019~\cite{liu2019ispd}, and ICCAD'2015~\cite{7372671} benchmarks demonstrate that \sys{} successfully inserts watermarks without compromising layout quality. Furthermore, we showcased \sys{}'s resiliency against watermark removal and forging attacks through comprehensive attack evaluations. 
 
\bibliographystyle{IEEEtran}
\bibliography{sample-base,bib}

\end{document}